\documentclass[aps, pra, twocolumn, superscriptaddress,floatfix]{revtex4-2}
\usepackage[]{physics}
\usepackage[]{amsmath}
\usepackage{amssymb}
\usepackage{mathrsfs}

\usepackage[]{graphicx}
\graphicspath{{src/figures/}}
\usepackage[]{mathtools}
\usepackage{esint}
\usepackage[normalem]{ulem}
\usepackage{cancel}
\usepackage[]{slashed}
\usepackage[colorlinks=true]{hyperref}
\usepackage[caption=false]{subfig}
\usepackage{cases}
\usepackage[]{comment}

\newcommand{\figref}[1]{\figurename~\ref{#1}}

\newcommand{\secref}[1]{Sec.~\ref{#1}}
\newcommand{\appref}[1]{Appendix~\ref{#1}}

\newcommand{\pos}[2]{\vb{#1} _ {#2}}

\newcommand{\omegap}{\omega _ \mathrm{p}}
\newcommand{\omegasp}{\omega _ \mathrm{sp}}

\newcommand{\omegaq}{\omega _ \mathrm{q}}

\newcommand{\xq}{\pos{x}{\mathrm{q}}}
\newcommand{\Rq}{\pos{r}{\mathrm{q}}}
\newcommand{\zq}{z _ \mathrm{q}}

\newcommand{\ux}{\pos{u}{x}}
\newcommand{\uy}{\pos{u}{y}}
\newcommand{\uz}{\pos{u}{z}}

\newcommand{\uR}{\vb{u} _ +}
\newcommand{\uL}{\vb{u} _ -}

\newcommand{\epsD}{\tensor{\epsilon} _ \mathrm{D}}
\newcommand{\epsEO}{\tensor{\epsilon} _ \mathrm{EO}}
\newcommand{\epsd}{\epsilon _ \mathrm{d}}
\newcommand{\epsg}{\epsilon _ \mathrm{g}}

\newcommand{\epsm}{\epsilon _ \mathrm{-}}
\newcommand{\epsEOp}{\epsilon _ \mathrm{EO,+}}
\newcommand{\epsEOm}{\epsilon _ \mathrm{EO,-}}
\newcommand{\epsEOpm}{\epsilon _ \mathrm{EO,\pm}}
\newcommand{\epsDx}{\epsilon _ {\mathrm{D},x}}
\newcommand{\epsDp}{\epsilon _ {\mathrm{D},+}}
\newcommand{\epsDm}{\epsilon _ {\mathrm{D},-}}
\newcommand{\epsDpm}{\epsilon _ {\mathrm{D},\pm}}
\newcommand{\epslss}{\tensor\epsilon'' _ <}
\newcommand{\epsgrt}{\tensor\epsilon'' _ >}

\newcommand{\varepsd}{\varepsilon _ \mathrm{d}}

\begin{document}

\title{
  Fluctuation-induced Hall-like lateral forces in a chiral-gain environment
}

\author{Daigo Oue}
\email{daigo.oue@gmail.com}
\affiliation{Instituto Superior T\'{e}cnico--University of Lisbon and Instituto de Telecomunica\c{c}\~{o}es, Lisbon 1049-001, Portugal}
\affiliation{The Blackett Laboratory, Imperial College London, London SW7 2AZ, United Kingdom}
\affiliation{RIKEN Centre for Advanced Photonics, Saitama 351-0198, Japan}
\author{M\'{a}rio G. Silveirinha}
\affiliation{Instituto Superior T\'{e}cnico--University of Lisbon and Instituto de Telecomunica\c{c}\~{o}es, Lisbon 1049-001, Portugal}

\date{\today}

\begin{abstract}
  Here, we demonstrate that vacuum fluctuations can induce lateral forces on a small particle positioned near a translation-invariant uniform non-Hermitian substrate with chiral gain.
  This type of non-Hermitian response can be engineered by biasing a low-symmetry conductor with a static electric field and is rooted in the quantum geometry of the material through the Berry curvature dipole.
  The chiral-gain material acts as an active medium for a particular circular polarisation handedness, while serving as a passive, dissipative medium for the other polarisation handedness.
  Owing to the nonreciprocity and gain characteristics, momentum is continuously exchanged in a preferred direction parallel to the surface between the test particle and the surrounding electromagnetic field, giving rise to lateral forces. 
  Interestingly, the force can be viewed as a fluctuation-induced drag analogous to a Hall force.
  Indeed, although the gain is driven by an electric current, the resulting force acts perpendicular to the bias---unlike conventional current-drag effects. This effect stems from the skewed propagation characteristics of surface modes and gain-momentum locking.  
  Our theory reveals a Hall-like asymmetry in the field correlations and establishes a novel link between quantum geometry and fluctuation-induced phenomena, offering new possibilities for nanoscale control via tailored electromagnetic environments. 
\end{abstract}

\maketitle

\section{Introduction}
\label{sec:intro}
Non-Hermitian systems have attracted the attention and curiosity of researchers and have been extensively studied across a broad range of fields, including optics and photonics, acoustics, electronic circuits and condensed matter physics \cite{el2018non,ashida2020non,bergholtz2021exceptional,ding2022non,okuma2023non}.
Non-Hermiticity---arising from loss and/or gain---may dramatically modify the system's response, giving rise to phenomena that have no counterparts in Hermitian platforms. 
These include exceptional points and related non-trivial topological structures.
The distinct characteristics of non-Hermitian systems have enabled a variety of novel effects in optics and photonics, including enhanced lasing, unidirectional transmission, and perfect absorption.
Moreover, in condensed matter physics, the non-Hermiticity gives rise to novel phase transitions, such as PT-symmetric phonon lasing~\cite{zhang2018phonon}, re-entrant superconductivity~\cite{yamamoto2019theory}, and non-Hermitian many-body localisation~\cite{hamazaki2019non}.

Over the past decade, it has been shown that non-Hermitian electromagnetic responses---such as optical gain---can be induced and controlled by driving nonequilibrium dynamics, enabling active tuning of material properties in optical and photonic systems.
For example, setting a lossy medium in motion at a constant speed can lead to optical gain~\cite{silveirinha2014optical,silveirinha2014spontaneous,lannebere2016wave,lannebere2016negative}, which has been identified as the origin of the Zeldovich superradiance~\cite{zel1971generation,zel1972amplification} and vacuum-fluctuation-induced noncontact frictional forces---quantum friction ~\cite{schaich1981dynamic,hoye1992friction,brevik1993friction,pendry1997shearing}.
Similarly, instead of physically moving the system, an electrostatic bias can be applied to induce electric carrier drift, leading to optical gain~\cite{morgado2017negative,morgado2018drift,morgado2020nonlocal,morgado2021active,morgado2022directional} and related phenomena, such as Coulomb drag~\cite{gramila1991mutual,persson1998theory,shapiro2010thermal,volokitin2011quantum,dean2016nonequilibrium,shapiro2017fluctuation,narozhny2016coulomb}, which parallels quantum friction.

Applying an electrostatic bias can also induce electro-optic effects, such as the Pockels and DC Kerr effects.
These phenomena manifest as changes in the refractive index (or more generally, the permittivity tensor) in response to an applied electric field.
Of particular relevance to this study, recently, it was theoretically shown that an electrostatic bias may give rise to gain terms in the dielectric response function~\cite{lannebere2022nonreciprocal,rappoport2023engineering,morgado2024non,fernandes2024exceptional}, a phenomenon referred to as the non-Hermitian electro-optic effect (NHEO).
This effect originates from the quantum geometry of low-symmetry materials, where a nonlinearity in the transport equation arises due to the coupling between the Berry curvature of Bloch electrons and an applied electric field.
Specifically, the combination of a static bias with Berry curvature dipoles in low-symmetry conductors---such as twisted bilayers~\cite{rappoport2023engineering} and trigonal tellurium~\cite{morgado2024non}---generally results in polarisation-dependent optical gain.
Furthermore, as the Berry curvature acts as an effective magnetic field in the presence of the electrostatic bias, electric carriers in such materials follow skewed trajectories under the applied bias, and the dielectric response tensor also gains a magneto-optical-like conservative component~\cite{morgado2024non}.
Thereby, reflecting the nonequilibrium nature due to the bias-induced carrier motion, the dielectric response of such systems acquires non-conservative and nonreciprocal components, and thus lacks both Hermitian and transpose symmetries (i.e., $\tensor{\epsilon} ^ \dagger \neq \tensor{\epsilon}$ and $\tensor{\epsilon} ^ \top \neq \tensor{\epsilon}$).
Note that the system can exhibit a nonreciprocal response, since the bias-induced carrier motion (i.e., electric current) is odd under time reversal.

Interestingly, for some material symmetry groups, the non-Hermitian response exhibits chiral properties, such that left- and right-handed circularly polarised fields experience different effects---for example, one handedness may be dissipated while the other is amplified.
It has been shown that such \textit{chiral-gain} responses can enable transistor-like distributed behaviour, which may be exploited for optical isolation, amplification~\cite{lannebere2022nonreciprocal,rappoport2023engineering} and terahertz lasing~\cite{hakimi2023chiral}.
The polarization dependence in these low-symmetry conductors can be understood as a consequence of the skewed trajectories followed by Bloch electrons.

In this work, we study fluctuation-induced phenomena in a chiral-gain environment.
A well-known example of such phenomena in passive systems is the Casimir effect---an attractive interaction between two plates, mediated by the fluctuation of electromagnetic fields in their surroundings~\cite{casimir1948attraction}.
The Casimir--Polder effect~\cite{casimir1948influence} describes a similar attractive force in a configuration involving an atom and a plate instead of two plates.

\begin{figure}[tbp]
  \centering
  \includegraphics[width=.7\linewidth]{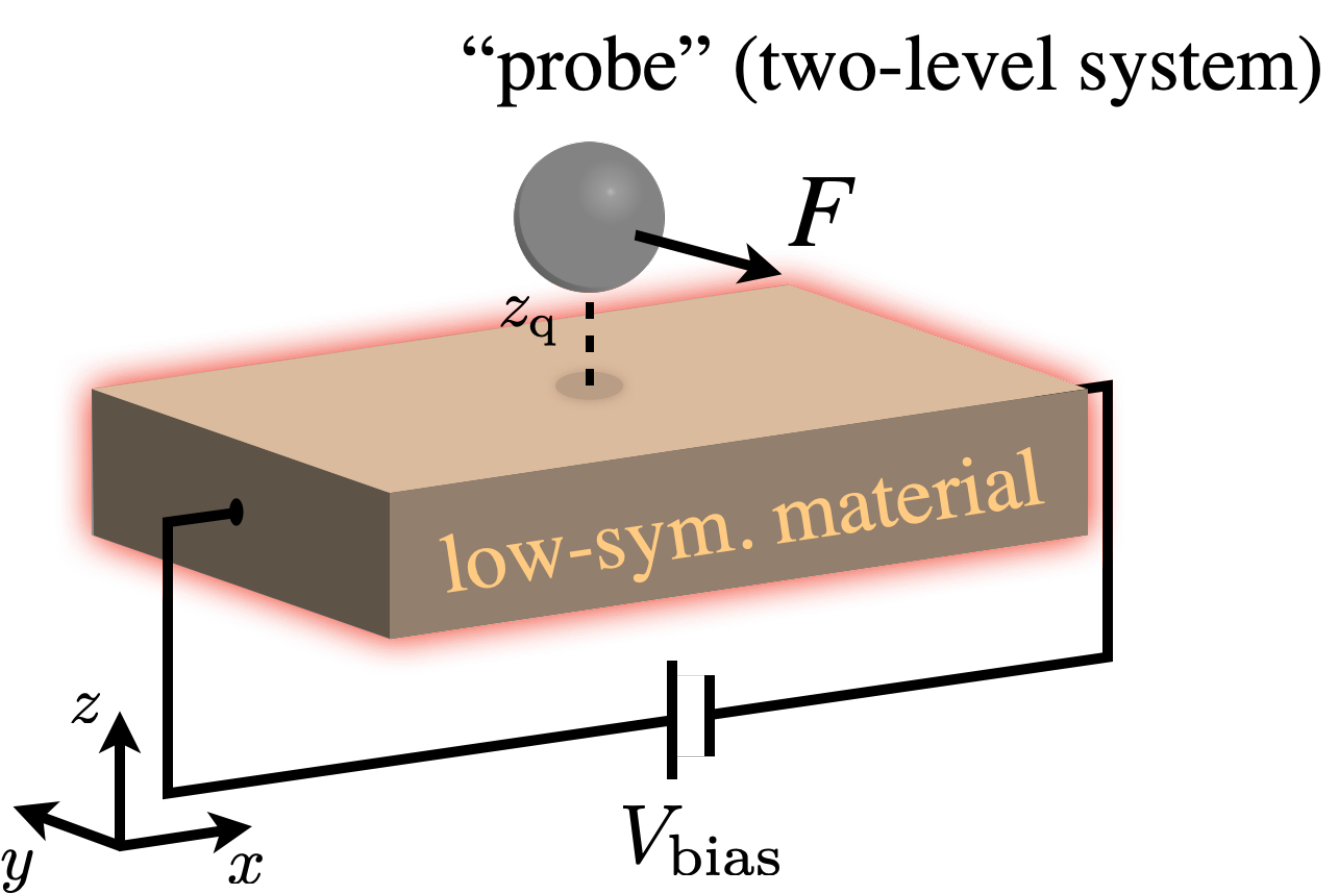}
  \caption{
    Schematic illustration of the setup under study.
    A point particle is placed at a distance $z _ \mathrm{q}$ above a low-symmetry material exhibiting a chiral-gain response.
    The electrostatic bias is applied along the $x$ direction.
    The particle is modelled as a two-level system.
  }
  \label{fig:fig1}
\end{figure}

Here, we consider the Casimir--Polder configuration in the presence of a chiral-gain medium as shown in \figref{fig:fig1}.
A small particle modelled by a two-level system is placed above the material substrate.
The chiral-gain medium occupies the lower half-space ($z < 0$), and the position of the particle is denoted by $\Rq = \xq + \zq \uz$, where $\xq := x _ \mathrm{q} \ux + y _ \mathrm{q} \uy$ represents the transverse part of the position vector of the particle.
The unit vectors along the $x$, $y$, and $z$ directions are denoted as $\vb{u} _ {x,y,z}$.

In our setup, the electric bias $E _ \mathrm{bias}$ is applied along the $x$ direction. Under an applied electric bias, the material response of conductors belonging to the 32 point group takes the form~\cite{morgado2024non,lannebere2025symmetry}: 
\begin{align}
  \tensor{\epsilon}(z < 0) = \epsD + \epsEO = \epsd I _ {3\times 3} + i \epsg \ux \times I _ {3\times3}.
  \label{eq:eps_NHEO}
\end{align}
The cross symbol ``$\times$'' represents the vector product operator, and $I _ {3 \times 3} := \ux\ux + \uy\uy + \uz\uz$ is the 3-by-3 identity matrix.
The diagonal contribution $\epsD$ is determined by a conventional Drude-type dispersion,
$\epsd = 1 - \omegap ^ 2/(\omega ^ 2 + i \omega \gamma),$
with $\omegap$ the plasma frequency and $\gamma$ the collision frequency.
The off-diagonal contribution $\epsEO$ arises due to the electro-optic effect and is determined by
\begin{align}
  \epsg = \frac{\omega _ 0 \gamma}{\omega} \qty(\frac{2}{\gamma} + \frac{1}{\gamma - i\omega}),
  \label{eq: epsg}
\end{align}
where $\omega _ 0 = 4\pi\alpha _ e e c D E _ \mathrm{bias}/(\hbar \gamma)$ is a cyclotron-type frequency characterising the strength of the static electric bias~\cite{morgado2024non}.
Here, $e$ is the electron charge, $c$ is the speed of light, $\alpha _ e \approx 1/137$ is the fine structure constant, and $D$ is the strength of the (dimensionless) Berry curvature dipole of the low-symmetry conductor.
For tellurium, its magnitude is on the order of $D \sim 10 ^ {-4}$ according to the first-principles calculations~\cite{tsirkin2018gyrotropic,morgado2024non}, though significantly larger values can be achieved in other materials from different symmetry groups~\cite{lannebere2025symmetry}.
In experimental studies (see, e.g., \cite{vorobev1979optical,shalygin2012current}), the kinetic Faraday rotation due to the gyrotropic response was measured, validating, in part, the result in Eq.~\eqref{eq:eps_NHEO}.

The particle is described as a two-level system whose transition frequency is denoted by $\omegaq$.
The corresponding Hamiltonian is
\begin{align}
  H _ \mathrm{q} = \frac{\hbar\omegaq}{2} \sigma _ z,
\end{align}
where we introduced the population inversion operator $\sigma _ z = \ketbra{1}{1} - \ketbra{0}{0}$.

The remainder of this article is organised as follows.
In \secref{sec:nheo}, we discuss the properties of the chiral-gain medium, particularly polarisation-dependent gain.
The field quantisation in the gain environment is discussed in \secref{sec:field_quantisation}. 
We describe how to express the fluctuating current operator $\vb{j} ^ {-}$ in terms of the non-Hermitian material response and provide the `bare' Hamiltonian for the field.
In addition, we characterise the field correlation functions, which play the central role in effectively describing the two-level system in the chiral-gain environment.
In \secref{sec:effectve_desc}, following the Lindblad formalism, we derive a reduced quantum master equation that governs the dynamics of the internal degrees of freedom of the two-level system in the gain environment.
Building on these results, in  \secref{sec:forces} we calculate the fluctuation-induced force acting on the two-level system and demonstrate that it includes a lateral component arising from the polarisation-dependent gain.
A discussion and the final conclusion are presented in \secref{sec:conclusion}.

\section{Chiral gain}
\label{sec:nheo}
As briefly discussed in Sec.~\ref{sec:intro}, previous studies have shown that the non-trivial Berry curvature dipole of certain low-symmetry conducting materials can induce distinctive electro-optic effects. These effects stem from the anomalous velocity of Bloch electrons, which includes a component governed by the Berry curvature, effectively acting as a magnetic field.
It has been shown that Berry curvature underlies a range of phenomena, including current-induced magnetisation~\cite{shalygin2012current,yoda2015current,furukawa2017observation,csahin2018pancharatnam,hara2020current,furukawa2021current}, tunable valley magnetisation~\cite{son2019strain,qin2021strain}, and circular photogalvanic effects~\cite{moore2010confinement,tsirkin2018gyrotropic,golub2020semiclassical,shi2023berry}, and nonlinear Hall effect~\cite{sodemann2015quantum,ma2019observation,du2021nonlinear}.
Moreover, the anomalous electron transport can render the dielectric tensor magneto-optical-like~\eqref{eq:eps_NHEO}~\cite{shalygin2012current,konig2019gyrotropic,shi2023berry,morgado2024non} leading to a nonreciprocal permittivity response ($\tensor\epsilon ^ \top \neq \tensor\epsilon$) that underpins the kinetic Faraday effect~\cite{vorobev1979optical, shalygin2012current}.

Additionally, as the electric bias drives the system into a nonequilibrium steady state characterised by a drift current, it becomes possible to extract energy from the moving carriers, resulting in optical gain.
Specifically, the bias introduces an additional component $\epsEO$ to the material response, enhancing its non-Hermitian character and potentially leading to optical gain ($\tensor\epsilon ^ \dagger \neq \tensor\epsilon$)~\cite{shi2023berry,morgado2024non}.

\subsection{Decomposition of the response tensor}
It is instructive to perform an eigen-decomposition of the electro-optic contribution $\epsEO$ to the dielectric tensor.
Since the matrix $\epsEO$ is normal [i.e, $\comm{\epsEO'}{\epsEO''} = 0$, where $\epsEO' := (\epsEO + \epsEO ^ \dagger)/2$ and $\epsEO'' := (\epsEO - \epsEO ^ \dagger)/(2i)$ are the Hermitian and non-Hermitian parts], it is unitary-diagonalisable (the eigenvectors of the matrix $\epsEO$ are mutually orthogonal),
\begin{subequations}
\begin{align}
  &\epsEO =
  \epsEOp \uR \uR ^ * + \epsEOm \uL \uL ^ *,
  \label{eq:eps_NHEO-decomposed}
  \\
  &\epsilon _ {\mathrm{EO},\pm} = \pm\epsg,\
  \\
  &\uR = \frac{\uy + i\uz}{\sqrt{2}},
  \quad
  \uL = \frac{\uy - i\uz}{\sqrt{2}},
\end{align}
\end{subequations}
where $\epsilon _ {\mathrm{EO},\alpha}$ is an eigenvalue, and $\vb{u} _ \alpha$ is the corresponding eigenvector ($\alpha = +, -$).

The non-Hermitian part of the matrix $\epsEO$ has the same structure as the full permittivity tensor [see Eq.~\eqref{eq:eps_NHEO-decomposed}] when it is eigen-decomposed,
\begin{align}
  &\epsEO'' = \frac{\epsEO - \epsEO ^ \dagger}{2i} =
  \epsEOp'' \uR \uR ^ * + \epsEOm'' \uL \uL ^ *,
  \label{eq:Im_eps_NHEO-decomposed}
\end{align}
Here, $\epsEOpm''$ denotes the imaginary part of $\epsEOpm$.
Note that the Drude contribution $\epsD''$ is already diagonal.
For convenience, we write
\begin{align}
    \epsD'' = \epsDx'' \ux\ux ^ * + \epsDp'' \uR\uR ^ * + \epsDm'' \uL\uL ^ *,
\end{align}
where $\epsDx'' = \epsDpm'' = \epsd''$.
It is also useful to write
\begin{align}
    \tensor\epsilon'' = 
    \sum _ {\ell = \mathrm{D}, \mathrm{EO}} \tensor\epsilon'' _ \ell =
    \sum _ {\ell} 
    \sum _ {\alpha}
    \epsilon _ {\ell,\alpha}''
    \vb{e} _ {\ell,\alpha}
    \vb{e} _ {\ell,\alpha} ^ *.
\end{align}
For passive systems, the matrix describing the non-conservative response $ \tensor\epsilon''$ must be positive definite to ensure that non-Hermitian light-matter interactions always result in material absorption.
Accordingly, the eigenvalues of $ \tensor\epsilon''$ are required to be positive.
In contrast, gain media can supply energy to the wave, allowing for gain interactions that may produce negative eigenvalues.
 
In the case under analysis, the eigenvalues $\epsilon'' _ {\ell,\alpha}$ determine whether the corresponding eigen-polarisation experiences dissipation ($\epsilon'' _ {\ell,\alpha} > 0$) or gain ($\epsilon'' _ {\ell,\alpha} < 0$) in the channel specified by $\ell$.
In the Drude channel ($\ell = \mathrm{D}$), all polarisations should experience dissipation as $\epsDx''=\epsDpm''=\epsd'' > 0$.
In contrast, in the electro-optic channel ($\ell = \mathrm{EO}$), the field may undergo either dissipation or gain:
the ``right-handed'' field experiences dissipation ($\epsEOp'' = +\epsg'' > 0$), while the ``left-handed'' field experiences gain ($\epsEOm'' = -\epsg'' < 0$).
Note that the terms ``right-handed'' and ``left-handed'' are defined with respect to the optical axis of the material ($+x$ axis).

Evidently, reversing the direction of the applied electric bias interchanges the circular polarisations that experience gain and dissipation.
We underline that the polarisations associated with gain and loss are determined by the material properties themselves; they are not dictated by the field distribution within the material or the direction of propagation.
The actual response of the material, whether dissipative or active, depends on the overlap between the field distribution and the eigen-polarisations that govern the non-Hermitian response. 

\begin{figure}[htbp]
  \centering
  \includegraphics[width=.8\linewidth]{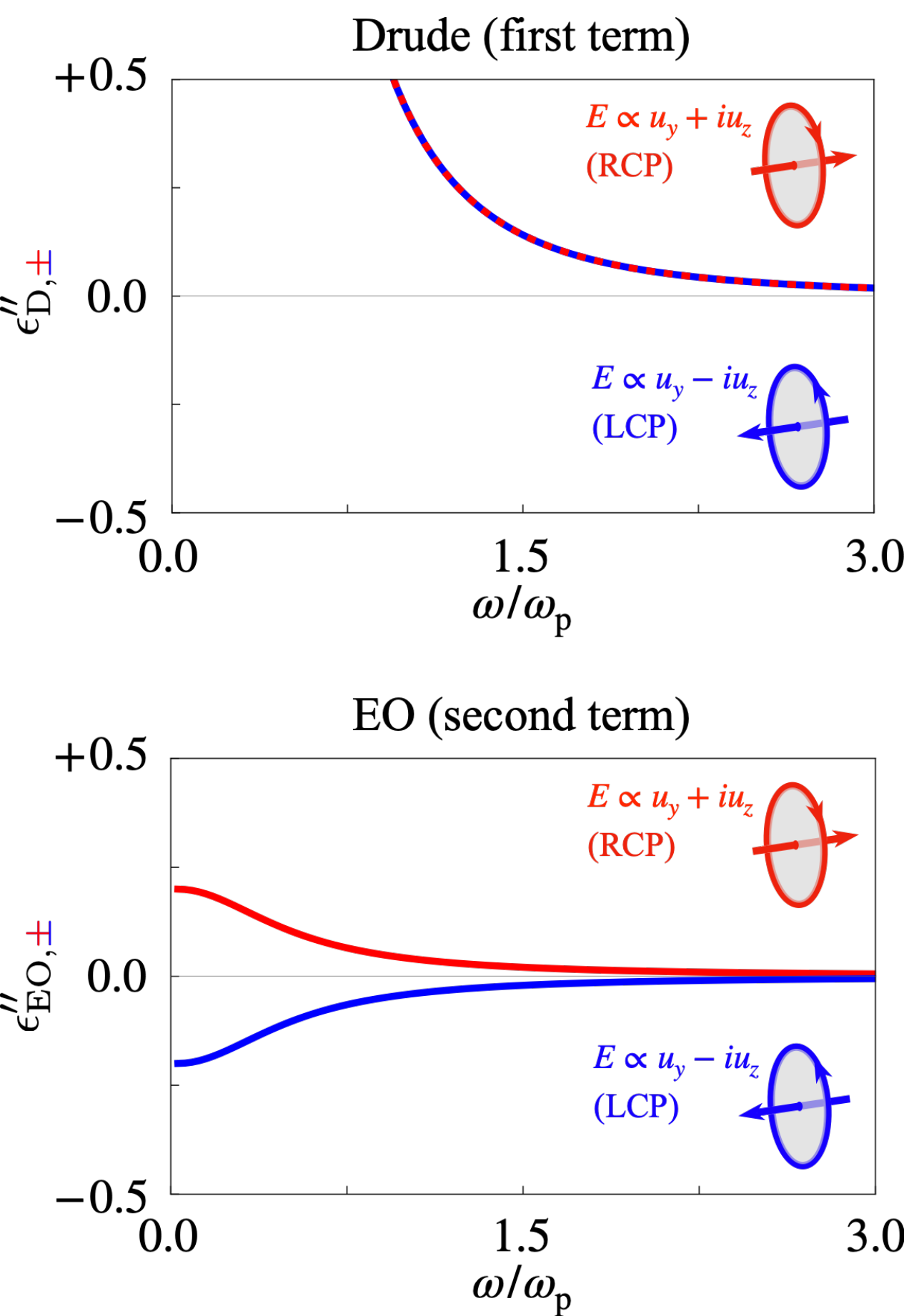}
  \caption{
    Two eigenvalues $\epsilon'' _ {\ell,\pm}$ of the non-Hermitian part of the dielectric response $\tensor\epsilon'' _ \ell$ in the Drude and electro-optic channels ($\ell = \mathrm{D}, \mathrm{EO}$) as a function of frequency.
    The dashed curve represents the response $\epsDpm''$ from the Drude channel, and the red and blue curves represent the one $\epsEOpm''$ from the electro-optic channel.
    In the Drude channel ($\ell = \mathrm{D}$), both the right-handed polarisation (RCP) and left-handed polarisation (LCP), $\uR$ and $\uL$, experience dissipation ($\epsDpm'' > 0$).
    On the other hand, in the electro-optic channel ($\ell = \mathrm{EO}$), the right-handed (left-handed) polarisation is subject to dissipation (gain) [$\epsEOp'' > 0$ ($\epsEOm'' < 0$)].
    The following parameters were used to generate the plot:
    $\gamma/\omegap = 0.5$ and $\omega _ 0/\omegap = 0.1$.
  }
  \label{fig:Im_eps}
\end{figure}

In \figref{fig:Im_eps}, we depict the frequency dependence of the eigenvalues $\epsilon'' _ {\ell,\pm}$ for the two distinct circular polarisations in each channel ($\ell = \mathrm{D}, \mathrm{EO}$).
Note that the non-Hermitian response, including the polarisation-dependent gain in the electro-optic channel, is naturally suppressed in the high-frequency limit, $\epsilon'' _ {\ell,\pm} \rightarrow 0$ ($\omega \rightarrow \infty$).
This is consistent with the fact that the polarisation-dependent gain is due to the motion of electric carriers within the medium, which cannot keep pace with a dynamic field oscillation when it becomes excessively rapid.
For sufficiently low frequencies ($\omega \approx 0$), the gain effect becomes less significant, as the dissipative Drude contribution $\epsd$ dominates the overall non-Hermitian electromagnetic response ($\abs{\epsg}/\abs{\epsd} \approx 3\omega_0\gamma/\omegap ^ 2 \ll 1$).

\subsection{Surface plasmon resonance}
\label{subsec:spr}
The surface plasmon resonance (SPR) is the frequency at which charge oscillations at a metal-dielectric interface couple strongly to an electromagnetic wave.
In the unbiased case, the SPR is isotropic: it occurs when $\epsd + 1 = 0$, giving $\omega_\mathrm{sp} \sim \omegap/\sqrt{2} \approx 0.7\omegap$, independent of the propagation direction.
When a chiral gain (electric bias) is introduced, this symmetry is broken.
The dielectric response becomes direction-dependent, similar to a magnetized passive plasma~\cite{silveirinha2018fluctuation,hassani2018spontaneous}.
The dispersion of the resulting surface plasmons follows from the poles of the reflection and transmission coefficients (see Appendix \ref{appx:transmission} for the derivation).

Within the quasi-static approximation, the poles can be found by solving
$
    (\epsd+1)\abs{\vb{k}}-\epsg k _ y = 0.
$
The solutions are:
\begin{align}
    \omega = -i\frac{\gamma}{2}+\frac{\omega _ 0 k _ y}{2\abs{\vb{k}}}
    \pm\sqrt{
    \omega _ \mathrm{sp} ^ 2 - \frac{\gamma ^ 2}{4} 
    +\qty(\frac{\omega _ 0 k _ y}{2\abs{\vb{k}}}) ^ 2
    +i\gamma\frac{\omega _ 0 k _ y}{\abs{\vb{k}}}
    }.
    \label{eq:omsp_mod}
\end{align}
If the electro-optic effect is sufficiently weak (i.e., $\omega _ 0$ is much smaller than any other relevant frequencies), we can approximate
\begin{align}
    \omega \approx 
    \pm\widetilde\omega _ \mathrm{sp}
    +\frac{\omega _ 0 k _ y}{2\abs{\vb{k}}}
    -i\frac{\gamma}{2}\qty(
    1-\frac{\omega _ 0 k _ y}{\widetilde\omega _ \mathrm{sp}\abs{\vb{k}}}
    ),
    \label{eq:omsp_mod_approx}
\end{align}
where we defined $\widetilde\omega _ \mathrm{sp} = \sqrt{\omega _ \mathrm{sp} ^ 2 - \gamma ^ 2/4}$.
The real part of the complex dispersion relation clearly exhibits the direction dependence of the SPR.
Remarkably, the SPP dispersion in Eq.~\eqref{eq:omsp_mod_approx} exhibits asymmetric behavior along the $y$-direction---that is, perpendicular to the applied bias.
In particular, the imaginary part, which governs the plasmon lifetime, is sensitive to the sign of $k_y$.
As we will further discuss in Section \ref{subsec:quasi-static}, this asymmetry underlies the emergence of a fluctuation-induced Hall-like lateral force.

The direction dependence of surface plasmons arises from the spin-momentum locking~\cite{bliokh2015quantum,van2016universal} combined with our polarisation-dependent gain, which originates a ``gain-momentum locking''~\cite{serra2024gain,prudencio2024topological}.
Specifically, since the handedness of the plasmons is determined by their propagation direction, and the non-conservative effects are governed by the handedness, the gain becomes locked to a particular propagation direction, while the opposite direction experiences enhanced dissipation.

The imaginary part of the complex dispersion relation determines the system's stability.
One can show that
\begin{align}
    \operatorname{Max}\Im\qty{\omega} =
    -\frac{\gamma}{2}\qty(
    1-\frac{\omega _ 0}{\widetilde\omega _ \mathrm{sp}}
    ) < 0.
\end{align}
Thus, the system is stable when the electro-optic effect is weak ($\omega _ 0 < \widetilde\omega _ \mathrm{sp}$).

\section{Field quantisation}
\label{sec:field_quantisation}
The objective of this article is to study fluctuation-induced forces when a particle is placed near the chiral-gain environment considered in the previous section.
To this end, next, we introduce the formalism required to characterise the quantised electromagnetic fields.

\subsection{Particle-field interaction}
The interaction between the particle and the surrounding electromagnetic environment is described within the dipole-coupling Hamiltonian (within the secular approximation),
\begin{align}
  H _ \mathrm{int} = -\int _ 0 ^ \infty \big(\vb{p} ^ - \cdot \vb{E} ^ + (\Rq,\omega) + \vb{p} ^ + \cdot \vb{E} ^ - (\Rq,\omega)\big) \dd{\omega},
  \label{eq:H_int}
\end{align}
where we introduced the electric field operators, $\vb{E} ^ -$ and $\vb{E} ^ + = \qty(\vb{E} ^ -) ^ \dagger$, and transition operators, $\vb{p} ^ - = \vb{d} _ \mathrm{e} \ketbra{0}{1}$ and $\vb{p} ^ + = \qty(\vb{p} ^ -) ^ \dagger$, with $\vb{d} _ \mathrm{e}$ the transition dipole moment of the two-level system.
Note that the frequency integration limits may be omitted for conciseness when appropriate in the following.
The electric field operator $\vb{E} ^ -$ has bosonic nature and satisfies an inhomogeneous wave equation, which is derived from Maxwell's equations,
\begin{align}
  \qty[\nabla \times \nabla \times - \frac{\omega ^ 2}{c ^ 2} \tensor\epsilon(\vb{r},\omega)]\vb{E} ^ -(\vb{r},\omega) = i\omega \mu _ 0 \vb{j} ^ -(\vb{r},\omega),
  \label{eq:maxwell}
\end{align}
where $\mu _ 0$ is the vacuum permeability, and the electric current source $\vb{j} ^ -$ represent fluctuations in our system, which we shall discuss in more detail in the subsequent sections.
Note that for conciseness, we shall omit position and/or frequency arguments in the following, whenever there is no risk of confusion.

Following the standard phenomenological quantisation procedure of macroscopic quantum optics~\cite{gruner1995correlation,gruner1996green,dung1998three,scheel1998qed,raabe2007unified}, we introduce the photonic Green's function $\tensor{G} := \qty[\nabla \times \nabla \times - \omega ^ 2\tensor\epsilon / c ^ 2] ^ {-1}$ to solve the wave equation \eqref{eq:maxwell}. Specifically, the electric field operator can be expressed as,
\begin{align}
  \vb{E} ^ -(\pos{r}{1},\omega) = i\omega \mu _ 0 \int \tensor{G}(\pos{r}{1},\pos{r}{2}) \cdot \vb{j} ^ -(\pos{r}{2},\omega)\dd{\pos{r}{2}},
  \label{eq:E=Gj}
\end{align}
In the following, we examine how the presence of chiral gain modifies the quantisation of the electromagnetic field.

\subsection{Hamiltonian and Field Correlations}

In general, the quantised electromagnetic field in a non-Hermitian environment is described by two sets of harmonic oscillators~\cite{jeffers1993quantum,jeffers1996canonical,matloob1997electromagnetic,gardiner2004quantum,raabe2008qed,amooghorban2011casimir}.
The Hamiltonian will be
\begin{align}
  H _ \mathrm{f} 
  &= \int _ 0 ^ \infty \sum _ {\ell, \alpha} \int \limits _ {\epsilon _ {\ell, \alpha} ''(\vb{r},\omega) > 0} 
  (+\hbar \omega) f _ {\ell, \alpha} ^ \dagger(\vb{r},\omega) f _ {\ell, \alpha}(\vb{r},\omega)
  \dd{\vb{r}}\dd{\omega} 
  \notag \\
  &+\int _ 0 ^ \infty \sum _ {\ell, \alpha} \int \limits _ {\epsilon _ {\ell, \alpha} ''(\vb{r},\omega) < 0}
  (-\hbar \omega) f _ {\ell, \alpha} ^ \dagger(\vb{r},\omega) f _ {\ell, \alpha}(\vb{r},\omega)
  \dd{\vb{r}}\dd{\omega}, 
  \label{eq:H_f}
\end{align}
where $\epsilon _ {\ell, \alpha}''(\vb{r},\omega)$ is an eigenvalue, labeled by $\alpha$, of the non-Hermitian part of the dielectric response from the channel labeled by $\ell$, $\tensor\epsilon'' _ \ell (\vb{r},\omega) = [\tensor\epsilon _ \ell (\vb{r},\omega) - \tensor\epsilon _ \ell  ^ \dagger (\vb{r},\omega)]/(2i)$, and $f _ {\ell, \alpha}(\vb{r},\omega)$ is a bosonic annihilation operator associated with the mode specified by its labels and arguments $\qty(\ell, \alpha,\vb{r},\omega)$.
It is crucial to introduce bosonic operators for each mode in each physically distinct channel, as clarified in our recent work~\cite{silveirinha2025chiral}.
This is because each channel may provide distinct fluctuations.
Note that in Eq.~\eqref{eq:H_f}, positive-frequency oscillators are assigned to modes experiencing loss ($\epsilon'' _ {\ell, \alpha} > 0$), whereas negative-frequency oscillators to modes experiencing gain ($\epsilon'' _ {\ell, \alpha} < 0$).
This is a consequence of the fact that the roles of annihilation and creation operators are swapped in the presence of field amplification~\cite{jeffers1993quantum,jeffers1996canonical,matloob1997electromagnetic,gardiner2004quantum}.
This prescription also generates the correct Maxwell equations via the Heisenberg equations of motion, which have been justified by the input-output theory~\cite{jeffers1993quantum,jeffers1996canonical,matloob1997electromagnetic} and the path-integral formalism~\cite{amooghorban2011casimir}. It correctly describes the impact of gain media on the Casimir force~\cite{sambale2009impact}, spontaneous emission in PT-symmetric setups~\cite{akbarzadeh2019spontaneous,ren2021quasinormal,franke2021fermi}, and quantum friction in which motion-induced gain plays a vital role~\cite{pendry1998can,silveirinha2014theory,silveirinha2014optical,silveirinha2014spontaneous,oue2024stable}.

The field quantisation can be completed by establishing an appropriate relation between the bosonic operators $f _ \alpha$ and the fluctuating current source $\vb{j} ^ -$.
The relation should be consistent with the fundamental canonical commutation relation of the field operator $\comm{\vb{E}(\pos{r}{1})}{\vb{B}(\pos{r}{2})} = i\hbar \partial _ {\pos{r}{1}}\times I _ {3\times3} \delta _ {\pos{r}{1},\pos{r}{2}}$, where we used a shorthand notation $\delta _ {\pos{r}{1},\pos{r}{2}} := \delta(\pos{r}{1}-\pos{r}{2})$.
In their pioneering works~\cite{gruner1995correlation,gruner1996green}, Gruner and Welsch demonstrated that setting
\begin{align}
  \vb{j} ^ -
  &= \frac{\omega}{c}
  \sqrt{\frac{\hbar}{\pi\mu _ 0}} \sqrt{\tensor\epsilon''} \cdot \vb{a},
  \label{eq:j-}
\end{align}
works if the system is purely dissipative, where all the eigenvalues of the non-Hermitian part of each response tensor are positive [i.e.~$\tensor\epsilon'' = \sum _ {\ell, \alpha} \epsilon'' _ {\ell, \alpha} \vb{e} _ {\ell, \alpha} \vb{e} _ {\ell, \alpha} ^ *$ such that $\epsilon'' _ {\ell, \alpha} > 0$ for all $(\ell,\alpha)$]. 
Note that we introduced $\vb{a} = \sum _ {\ell, \alpha} \vb{e} _ {\ell, \alpha} f _ {\ell, \alpha}$, and the square root can be written in terms of the eigendecomposition,
$\sqrt{\tensor\epsilon''} = \sum _ {\ell, \alpha} \sqrt{\epsilon'' _ {\ell, \alpha}} \vb{e} _ {\ell, \alpha} \vb{e} _ {\ell, \alpha} ^ *$.
With Eq.~\eqref{eq:j-}, we can readily evaluate the symmetrised current correlation function,
\begin{align}
  \expval{\acomm{\vb{j} ^ -(\pos{r}{1})}{\vb{j} ^ +(\pos{r}{2})}} = 
  \frac{\hbar}{\pi\mu _ 0}
  \frac{\omega ^ 2}{c ^ 2}
  \tensor\epsilon''(\pos{r}{1}) \delta _ {\pos{r}{1},\pos{r}{2}},
  \label{eq:jj}
\end{align}
where we have defined $\vb{j} ^ + = (\vb{j} ^ -) ^ \dagger$.
The anticommutation relation is defined as $\acomm{\vb{A}}{\vb{C}} = \vb{A}\vb{C} + \qty(\vb{C}\vb{A}) ^ \top$.
Equation \eqref{eq:jj} can be viewed as a fluctuation-dissipation relation in that the current-current correlation is connected to the non-Hermitian part of the response.
However, setting the fluctuating current operator as in Eq.~\eqref{eq:j-} does not work in the presence of gain: 
The non-conservative part $\tensor\epsilon''$ of the response is no longer positive definite, but the correlation function should be positive definite by definition so that Eq.~\eqref{eq:jj} becomes inapplicable.
The fundamental reason for this breakdown is that the square-root decomposition of the matrix, $\tensor\epsilon'' = \sqrt{\tensor\epsilon''} \cdot \sqrt{\tensor\epsilon''} ^ \dagger$, is no longer valid in the presence of gain~\footnote{%
  For any $\vb{E}$, we have $\vb{E} ^ * \cdot \sqrt{\tensor\epsilon''} \cdot \sqrt{\tensor\epsilon''} ^ \dagger \cdot \vb{E} = \abs{\sqrt{\tensor\epsilon''} ^ \dagger \cdot \vb{E}} ^ 2 > 0$ so that $\sqrt{\tensor\epsilon''} \cdot \sqrt{\tensor\epsilon''} ^ \dagger$ is positive definite, while $\tensor\epsilon''$ is not.%
}, and Eq.~\eqref{eq:j-} becomes ill-defined.

To address this issue, we shall proceed with a generalised prescription~\cite{scheel1998qed,raabe2008qed}, which remains valid even in the presence of gain.
In Refs.~\cite{scheel1998qed,raabe2008qed}, it was proved that taking the absolute value of the non-conservative part of the permittivity and swapping the roles of annihilation and creation operators does the trick,
\begin{align}
  \vb{j} ^ -
  &= \frac{\omega}{c}
  \sqrt{\frac{\hbar}{\pi\mu _ 0}} \sqrt{\abs{\tensor\epsilon''}} \cdot (\vb{a} + \vb{b} ^ \dagger),
  \label{eq:j-gen}
\end{align}
where the square root is defined in terms of the channel-wise eigen-decomposition,
$
\sqrt{\abs{\tensor\epsilon''}} = \sum _ {\ell, \alpha} \sqrt{\abs{\epsilon _ {\ell, \alpha}''}} \vb{e} _ {\ell, \alpha} \vb{e} _ {\ell, \alpha} ^ *,
$
and the vector-valued operators are defined as
\begin{align}
  \vb{a} = \sum _ {\epsilon'' _ {\ell, \alpha} > 0} \vb{e} _ {\ell, \alpha} f _ {\ell, \alpha},
  \quad
  \vb{b} ^ \dagger = \sum _ {\epsilon'' _ {\ell, \alpha} < 0} \vb{e} _ {\ell, \alpha} f _ {\ell, \alpha} ^ \dagger.
\end{align}
With these vector-valued operators, the Hamiltonian~\eqref{eq:H_f} can be written in a compact form,
\begin{align}
  H _ \mathrm{f}
  = \int 
  \hbar \omega 
  \qty[
  \vb{a} ^ \dagger(\vb{r},\omega) \cdot \vb{a}(\vb{r},\omega)
  -\vb{b} ^ \dagger(\vb{r},\omega) \cdot \vb{b}(\vb{r},\omega)
  ]
  \dd{\vb{r}}\dd{\omega}.
  \label{eq:H_f-compact}
\end{align}
The generalised prescription~\eqref{eq:j-gen} reproduces Eq.~\eqref{eq:j-} in the absence of gain.
Moreover, we obtain a modified fluctuation-dissipation relation,
\begin{align}
  \expval{\acomm{\vb{j} ^ -(\pos{r}{1})}{\vb{j} ^ +(\pos{r}{2})}} 
  &= \frac{\hbar}{\pi\mu _ 0} \frac{\omega ^ 2}{c ^ 2}
  \abs{\tensor\epsilon''(\pos{r}{1})} \delta _ {\pos{r}{1},\pos{r}{2}}.
  \label{eq:jj-gen}
\end{align}
Note that the matrix $\abs{\tensor\epsilon''}$ on the right-hand side is positive definite, as it should be.
It is also worth noting that
\begin{align}
  &\expval{\vb{j} ^ -(\pos{r}{1})\vb{j} ^ +(\pos{r}{2})} 
  = \frac{\hbar}{\pi\mu _ 0} \frac{\omega ^ 2}{c ^ 2}
  \abs{\tensor\epsilon'' _ >(\pos{r}{1})} \delta _ {\pos{r}{1},\pos{r}{2}},
  \label{eq:j-j+}
  \\
  &\expval{\vb{j} ^ +(\pos{r}{2})\vb{j} ^ -(\pos{r}{1})} ^ \top 
  = \frac{\hbar}{\pi\mu _ 0} \frac{\omega ^ 2}{c ^ 2}
  \abs{\tensor\epsilon'' _ <(\pos{r}{1})} \delta _ {\pos{r}{1},\pos{r}{2}},
  \label{eq:j+j-}
\end{align}
where the positive (negative) component of the non-Hermitian part of the response tensor is defined as
\begin{align}
  \tensor\epsilon'' _ {>(<)} = \sum _ {\substack{\epsilon'' _ {\ell, \alpha} > 0 \\ (\epsilon'' _ {\ell, \alpha} < 0)}}\epsilon'' _ {\ell, \alpha} \vb{e} _ {\ell, \alpha} \vb{e} _ {\ell, \alpha} ^ *.
\end{align}
The ``absolute values'' are defined as $\abs{\tensor\epsilon'' _ {>,<}} = \sqrt{\qty(\tensor\epsilon'' _ {>,<}) ^ 2}$. Note that the Drude term of the permittivity only contributes to the dissipative part of the decomposition $\tensor\epsilon'' _ {>}$, whereas the electro-optic term has polarisation dependent contributions to both the dissipative ($\tensor\epsilon'' _ {>}$) and gain ($\tensor\epsilon'' _ {<}$) parts.

As the electric field operator is written in terms of the fluctuating current operator and of the system Green's function~\eqref{eq:E=Gj}, we can readily determine the field correlation functions from Eqs.~\eqref{eq:j-j+} and \eqref{eq:j+j-}.
There are two distinct field correlation functions,
\begin{align}
  &\expval{\vb{E} ^ - (\pos{r}{1}) \vb{E} ^ +(\pos{r}{2})} 
  = \frac{\hbar}{\pi\epsilon _ 0} \tensor{\overline{G}}(\pos{r}{1},\slashed{\vb{r}}) \cdot \abs{\epsgrt(\slashed{\vb{r}})} \cdot \tensor{\overline{G}} ^ \dagger(\pos{r}{2},\slashed{\vb{r}}),
  \label{eq:E-E+}
  \\
  &\expval{\vb{E} ^ + (\pos{r}{2}) \vb{E} ^ -(\pos{r}{1})} ^ \top 
  = \frac{\hbar}{\pi\epsilon _ 0}
  \tensor{\overline{G}}(\pos{r}{1},\slashed{\vb{r}}) \cdot \abs{\epslss(\slashed{\vb{r}})} \cdot \tensor{\overline{G}} ^ \dagger(\pos{r}{2},\slashed{\vb{r}}) 
  \label{eq:E+E-}
\end{align}
where we defined $\tensor{\overline{G}} = (\omega ^ 2/c ^ 2)G$.
The integration over the slashed variable $\slashed{\vb{r}}$ is implicit.
As discussed above, the non-Hermitian part $\epsilon''$ of the dielectric tensor characterises the fluctuations in the system, whereas the Green's function $\tensor{\overline{G}}$ is responsible for propagating the fluctuations from one point to another.
From Eqs. \eqref{eq:E-E+} and \eqref{eq:E+E-}, it is evident that both correlation functions are positive definite as it should be.
We can recognise that the order of $\pos{r}{1}$ and $\pos{r}{2}$ appearing on the right-hand side is different for Eqs.~\eqref{eq:E-E+} and \eqref{eq:E+E-}.
In addition, one can see that the loss contribution $\abs{\epsgrt}$ appears in the first correlation function~\eqref{eq:E-E+} while the gain counterpart $\abs{\epslss}$ appears in the second one~\eqref{eq:E+E-}.
This observation suggests that the two correlation functions represent ``inverse'' processes.
Indeed, it is well-known from perturbation theory (Fermi's golden rule) that the first one $\expval{\vb{E} ^ - \vb{E} ^ +}$ characterises the spontaneous decay rate~\cite{novotny2012principles}, whereas the second one $\expval{\vb{E} ^ + \vb{E} ^ -}$ controls the rate of the inverse process (i.e.~photon absorption by the system), as we shall see in the following section.

\subsection{Quasi-static approximation}
\label{subsec:quasi-static}

Next, we calculate the field correlation functions [Eqs.~\eqref{eq:E-E+} and \eqref{eq:E+E-}] for our setup, using a quasi-static approximation.
We note that the non-Hermitian part $\epsilon''$ vanishes for the upper-half space ($z > 0$) in the air region; therefore, we can focus on the lower-half space ($z < 0$) when performing the spatial integration.

In \appref{appx:green}, we derive an explicit semi-analytical formula for the Green's function, neglecting the effects of time retardation.
The Green's function is calculated in the spectral domain (denoted by $
{\bf{k}}$), corresponding to a Fourier transformation in the $x$ and $y$ directions, exploiting the translational symmetry of the system. It is given by:
\begin{align}
  \tensor{\overline{G}} _ {\vb{k}}(z _ 1, z _ 2)
  = -\frac{t _ {\vb{k}}/\epsd}{2\abs{\vb{k}}} \vb{k} _ + \vb{k} _ + e ^ {-\abs{\vb{k}}(z _ 1 - z _ 2)}
  \quad 
  (z _ 2 < 0 < z _ 1),
\end{align}
where we defined $\vb{k} _ + = \vb{k} + i\abs{\vb{k}}\uz$ with $\vb{k} = k _ x \ux + k _ y \uy$ and the transmission coefficient $t _ {\vb{k}} = 2\epsd/(\epsd - \epsg k _ y/\abs{\vb{k}} + 1)$, which is derived in Appendix~\ref{appx:transmission}.
The Green's function in the spatial domain is obtained through an inverse Fourier transform:
\begin{align}
  \tensor{\overline{G}}(\pos{r}{1},\pos{r}{2}) = \int \tensor{\overline{G}} _ {\vb{k}}(z _ 1,z _ 2) e ^ {i\vb{k}\cdot(\pos{x}{1} - \pos{x}{2})}\dd{\vb{k}},
\end{align}
where $\dd{\vb{k}} := \dd{k _ x}\dd{k _ y}/(2\pi) ^ 2$.
At a given position $z$, we can evaluate the first correlator~\eqref{eq:E-E+} in the spectral domain as
\begin{align}
  &\tensor{\gamma} _ \mathrm{L}(\vb{k},z) 
  = \int
  \expval{\vb{E} ^ - (\pos{x}{1},z) \vb{E} ^ +(\pos{x}{2},z)} e ^ {-i\vb{k} \cdot \pos{x}{12}}
  \dd{\pos{x}{12}} \nonumber
  \label{eq:gamma _ L}
  \\
  &=\frac{\hbar\abs{\vb{k}}}{\pi \epsilon _ 0}
  \abs{
    \frac{t _ {\vb{k}} e ^ {-\abs{\vb{k}}z}}{\epsd}
  } ^ 2
  \frac{\vb{k} _ +}{2\abs{\vb{k}}}
  \frac{\vb{k} _ + \cdot \abs{\epsgrt} \cdot \vb{k} _ + ^ *}{2\abs{\vb{k}} ^ 2}
  \frac{\vb{k} _ + ^ *}{2\abs{\vb{k}}},
\end{align}
where we separated the transverse and $z$ components of the position vector, $\vb{E} ^ \pm (\pos{x}{1},z) := \vb{E} ^ \pm (\pos{r}{1})| _ {z _ 1 = z}$.
Note that we applied the Fourier transformation to the transverse coordinate $\pos{x}{12} := \pos{x}{1} - \pos{x}{2}$.
In the lower-half space ($z < 0$), the dissipative part of the non-Hermitian response $\epsgrt$ can be explicitly written as follows:
\begin{align}
  &\epsgrt = 
    \epsD'' + \epsg'' \uR\uR ^ *.
\end{align}

\begin{figure*}[tbp]
  \centering
  \includegraphics[width=\linewidth]{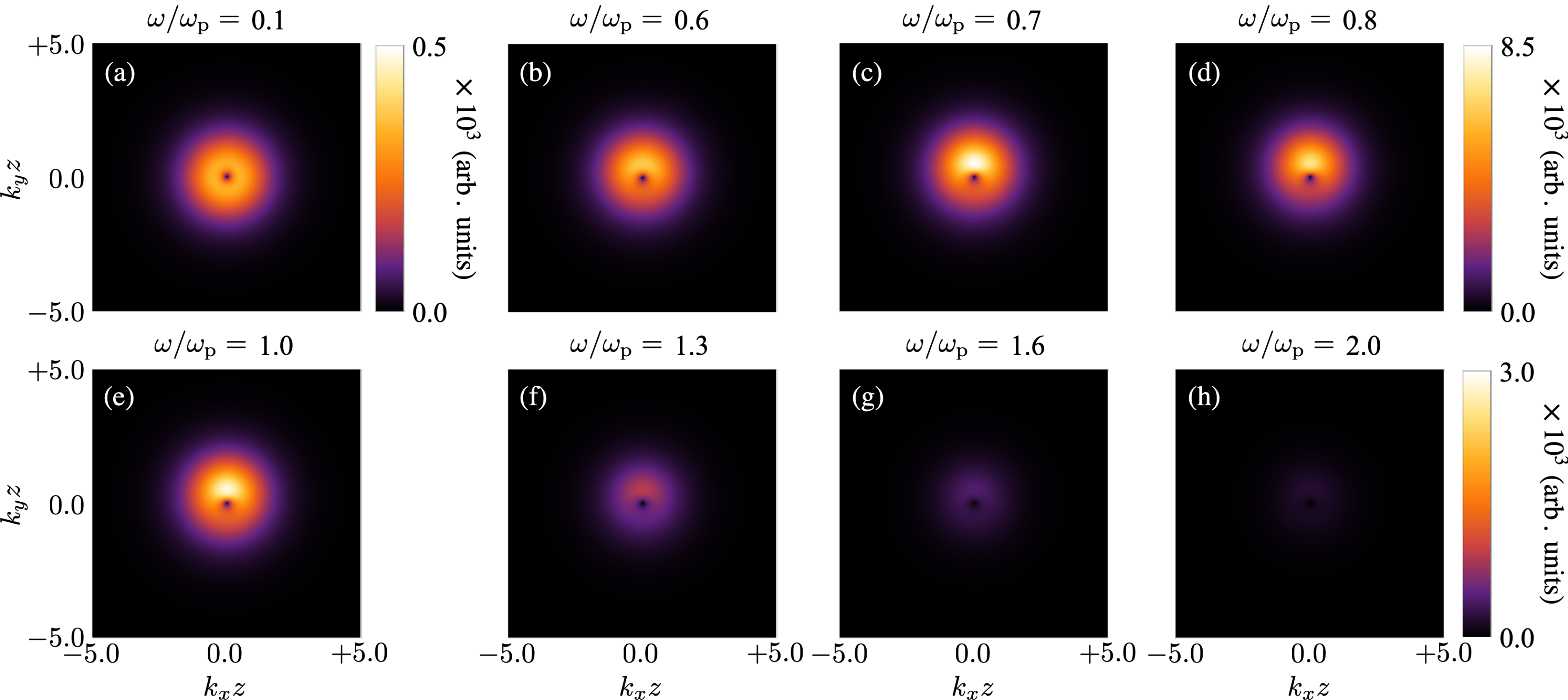}
  \caption{
    Amplitude of the emission rate spectrum $\norm{\tensor{\gamma} _ \mathrm{L}}$, characterising the directionality of the photon emission for various frequencies.
    The simulation parameters are:
    $\gamma/\omegap = 0.5$,
    $\omega _ 0/\omegap = 0.1$,
    and 
    $\omegap z/c = 0.0001$.
    Note that panel (a) has an individual colour bar;
    panels (b–d) share a colour bar located in the top right corner;
    and the colour bar for panels (e–h) is shown in the 
    bottom right corner.
  }
  \label{fig:E-E+}
\end{figure*}

Figure \ref{fig:E-E+} represents the spectral amplitude (trace norm) $\norm{\tensor{\gamma} _ {\mathrm{L}}(\vb{k},z)}$ of the first correlation function.
For low frequencies (\figref{fig:E-E+}a), the spectrum is isotropic so that there is no preferred direction for the radiative emission of photons.
This is consistent with the fact that the NHEO effect is less relevant at low frequencies, where the response is dominated by the isotropic dissipative Drude contribution.
On the other hand, at moderately higher frequencies, the spectrum becomes directional, exhibiting pronounced asymmetry between $+k_y$ and $-k_y$, in agreement with the anisotropic electromagnetic response of the chiral-gain medium.
In this case, the photon emission can be strongly directional.

The spectral asymmetry originates from the directional dependence of the SPR induced by the applied electric bias, as discussed in Subsection \ref{subsec:spr}.
Note that in the absence of the static bias, the resonance occurs for $\omegasp \approx \omegap/\sqrt{2}$, which is independent of the propagation direction. 
Accordingly, under a static bias, the $\tensor{\gamma} _ {\mathrm{L}}$ spectrum asymmetry is especially pronounced near $\omega = \omegap/\sqrt{2}$.
Thus, the photon emission from an excited two-level system may be lopsided, with a tendency towards the negative (positive) $y$ direction.
Note that the colour map darkens progressively as the frequency increases (compare, e.g., FIGs.~\ref{fig:E-E+}e-h). 

Following the same procedure, we can evaluate the spectrum of the second correlator~\eqref{eq:E+E-},
\begin{align}
  &\tensor{\gamma} _ \mathrm{G}(\vb{k},z) 
  = \int
  \expval{\vb{E} ^ + (\pos{x}{2},z) \vb{E} ^ -(\pos{x}{1},z)} ^ \top
  e ^ {-i\vb{k} \cdot \pos{x}{12}}
  \dd{\pos{x}{12}} \nonumber
  \label{eq:gamma _ G}
  \\
  &=\frac{\hbar\abs{\vb{k}}}{\pi \epsilon _ 0}
  \abs{
    \frac{t _ {\vb{k}} e ^ {-\abs{\vb{k}}z}}{\epsd}
  } ^ 2
  \frac{\vb{k} _ +}{2\abs{\vb{k}}}
  \frac{\vb{k} _ + \cdot \abs{\epslss} \cdot \vb{k} _ + ^ *}{2\abs{\vb{k}} ^ 2}
  \frac{\vb{k} _ + ^ *}{2\abs{\vb{k}}},
\end{align}
where the gain part of the non-Hermitian material response is $\epslss$ is
\begin{align}
  &\epslss = \epsm'' \uL\uL ^ *.
\end{align}

\begin{figure*}[tbp]
  \centering
  \includegraphics[width=\linewidth]{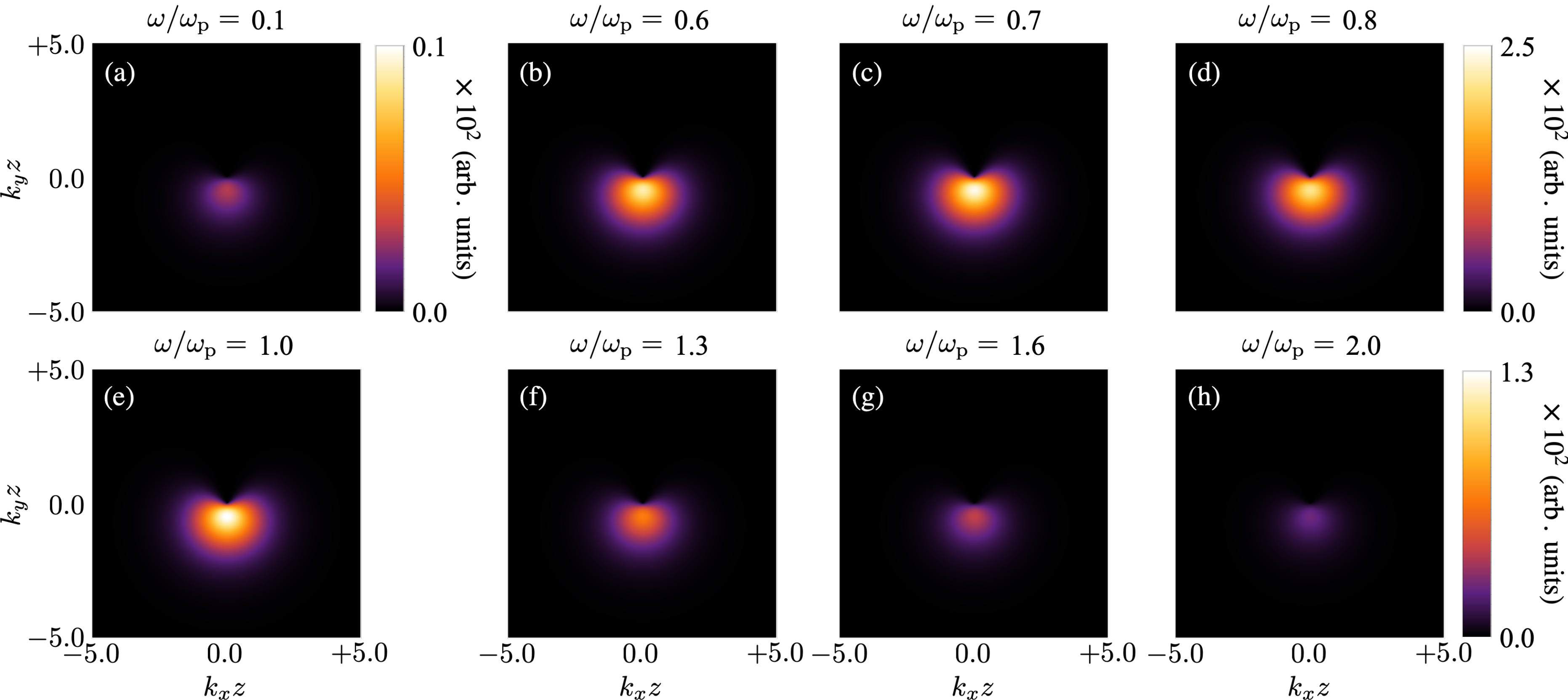}
  \caption{
    Amplitude of the absorption rate spectrum $\norm{\tensor{\gamma} _ \mathrm{G}}$, characterising the photon absorption for various frequencies.
    The simulation parameters are as in Fig.~\ref{fig:E-E+}.
    Note that panel (a) has an individual colour bar;
    panels (b–d) share a colour bar located in the top right corner;
    and the colour bar for panels (e–h) is shown in the 
    bottom right corner.
  }
  \label{fig:E+E-}
\end{figure*}

In \figref{fig:E+E-}, we depict the spectrum $\norm{\tensor{\gamma} _ {\mathrm{G}}(\vb{k},z)}$ of the second correlation function.
It is evident that spectral peaks appear on the negative side of $k _ y$.
This means that a two-level system can efficiently absorb photons propagating in the $-y$ direction from the environment.
Similar to $\tensor{\gamma} _ {\mathrm{L}}$, here the density plot also darkens as the high-frequency regime is approached. 
These observations are consistent with the fact that the chiral gain is weak compared to the dissipative Drude term at low frequencies, becomes most significant at intermediate frequencies---on the order of the collision frequency---, and is gradually turned off as we approach the high-frequency limit.
Note that the qualitative difference between the loss and gain spectra arises because the loss spectrum contains contributions from both the Drude term and the electro-optic term, whereas the gain spectrum is governed solely by the electro-optic term.

\section{Reduced master equation and steady-state}
\label{sec:effectve_desc}

Following a standard procedure for open quantum systems~\cite{breuer2002theory,gardiner2004quantum,rivas2012open}, next we develop an effective description of the two-level system, namely we obtain the evolution equation of the reduced density matrix, $\rho _ \mathrm{q} := \tr _ \mathrm{f} (\rho)$.
Here, $\tr _ \mathrm{f}$ represents the partial trace of the total density matrix over the electromagnetic field degrees of freedom.

Since the field operator behaves like a harmonic oscillator and evolves as $\vb{E} ^ - (\vb{r},\omega;t) \sim e ^ {-i\omega t}$ in the interaction picture, it is convenient to introduce
\begin{align}
  \vb{E} ^ -(\Rq,t) = \int \vb{E} ^ - (\Rq,\omega) e ^ {-i\omega t} \dd{\omega}.
  \label{eq:E-(t)}
\end{align}
As we are going to focus on the field evaluated at the position $\Rq$ of the particle, the position argument may be suppressed in this section for conciseness [e.g.~$\vb{E} ^ -(t) = \vb{E} ^ -(\Rq,t)$].

We start with the evolution equation for the total system (the integral form of the von Neumann equation),
$\rho(t) = \rho(0) + (i\hbar) ^ {-1} \int _ 0 ^ t \comm{H _ \mathrm{int}(t-s)}{\rho(t-s)}\dd{s}.$
Up to the second order in the interaction Hamiltonian~\eqref{eq:H_int}, we can write
\begin{align}
  \dv{\rho _ \mathrm{q}}{t} = -\frac{1}{\hbar ^ 2}\int _ {0} ^ \infty 
  \tr _ \mathrm{f} \comm{H _ \mathrm{int}(t)}{\comm{H _ \mathrm{int}(t-s)}{\rho(t)}}
  \dd{s},
  \label{eq:neumann_born_markov}
\end{align}
where we have assumed $\comm{H _ \mathrm{int}(t)}{\rho(0)} = 0$ and also adopted the Markov approximation, $\rho(s) \rightarrow \rho(t)$, letting the upper limit of the integral go to the infinity (i.e., $\int _ 0 ^ t \rightarrow \int _ 0 ^ \infty$).
We assume the electromagnetic field state approximately stays the same upon the evolution of the two-level system, factorising $\rho(t) \rightarrow \rho _ \mathrm{q}(t) \otimes \rho _ \mathrm{f}$.
This is nothing but the Born approximation.
The justification for employing the Born approximation, much like in open-systems theory, is the stability of the electromagnetic environment around the equilibrium point established by the DC bias.

In the following, we will assume that the electromagnetic subsystem stays in the vacuum state, $\rho _ \mathrm{f} = \ketbra{0}$.
The temperature effect may be taken into account by modifying the density matrix~\cite{silveirinha2025chiral}.
Then, the integrand in Eq.~\eqref{eq:neumann_born_markov} becomes
\begin{align}
  &\tr _ \mathrm{f} \comm{H _ \mathrm{int}(t)}{\comm{H _ \mathrm{int}(t-s)}{\rho(t)}}\notag \\
  &=\tr _ \mathrm{f} \comm{\vb{p} ^ +(t) \cdot\vb{E} ^ -(t)}{\comm{\vb{p} ^ -(t-s) \cdot\vb{E} ^ +(t-s)}{\rho _ \mathrm{q}(t)\otimes\rho _ \mathrm{f}}}
  \notag \\
  &+\tr _ \mathrm{f}
  \comm{\vb{p} ^ -(t) \cdot\vb{E} ^ +(t)}{\comm{\vb{p} ^ +(t-s) \cdot\vb{E} ^ -(t-s)}{\rho _ \mathrm{q}(t)\otimes\rho _ \mathrm{f}}},
  \label{eq:nested_comm}
\end{align}
where we have retained only the relevant terms: the conjugate pair of the field operators [e.g., $\vb{E} ^ -(t)$ and $\vb{E} ^ +(t-s)$] should be picked within the nested commutators to yield a finite contribution after the partial trace is taken.
If an irrelevant pair [e.g, $\vb{E} ^ -(t)$ and $\vb{E} ^ -(t-s)$] is chosen, it will eventually result in $\tr _ \mathrm{f} \qty{\rho _ \mathrm{f} \vb{E} ^ -(t)\vb{E} ^ -(t-s)} = 0$, and therefore, no contribution to the integral.
Expanding the nested commutators and using $\vb{p} ^ \pm(t) = \vb{p} ^ \pm e ^ {\pm i\omegaq t}$, the contribution from the first term in Eq.~\eqref{eq:nested_comm} can be written as
\begin{align}
  \frac{1}{2i}\Big(
  &\rho _ \mathrm{q} \vb{p} ^ - \cdot \tensor{\mathcal{C}} _ \mathrm{G} ^ \top \cdot \vb{p} ^ +
  +\vb{p} ^ + \cdot \tensor{\mathcal{C}} _ \mathrm{L} \cdot \vb{p} ^ - \rho _ \mathrm{q} 
  \notag\\
  &-(\tensor{\mathcal{C}} _ \mathrm{G} ^ \top \cdot \vb{p} ^ +) \cdot \rho _ \mathrm{q} \vb{p} ^ -
  -(\tensor{\mathcal{C}} _ \mathrm{L} \cdot \vb{p} ^ -) \cdot \rho _ \mathrm{q} \vb{p} ^ +
  \Big),
  \label{eq:integrand_1}
\end{align}
and the second term in Eq.~\eqref{eq:nested_comm} yields similar but conjugate results,
\begin{align}
  -\frac{1}{2i}\Big(
  &\vb{p} ^ - \cdot \tensor{\mathcal{C}} _ \mathrm{G} ^ * \cdot \vb{p} ^ + \rho _ \mathrm{q}
  +\rho _ \mathrm{q} \vb{p} ^ + \cdot \tensor{\mathcal{C}} _ \mathrm{L} ^\dagger \cdot \vb{p} ^ - 
  \notag\\
  &-(\tensor{\mathcal{C}} _ \mathrm{G} ^ * \cdot \vb{p} ^ +) \cdot \rho _ \mathrm{q} \vb{p} ^ -
  -(\tensor{\mathcal{C}} _ \mathrm{L} ^\dagger \cdot \vb{p} ^ -) \cdot \rho _ \mathrm{q} \vb{p} ^ +
  \Big).
  \label{eq:integrand_2}
\end{align}
Here, we have introduced
\begin{align}
  &\tensor{\mathcal{C}} _ \mathrm{L} = 
  2i\int _ 0 ^ \infty 
  \expval{\vb{E} ^ -(t) \vb{E} ^ +(t-s)}
  e ^ {+i\omegaq s}\dd{s},
  \label{eq:calG _ L}\\
  &\tensor{\mathcal{C}} _ \mathrm{G} = 
  2i\int _ 0 ^ \infty 
  \expval{\vb{E} ^ +(t-s) \vb{E} ^ -(t)} ^ \top
  e ^ {+i\omegaq s}\dd{s},
  \label{eq:calG _ G}
\end{align}
where we have written $\expval{\ldots} = \tr _ \mathrm{f}(\rho _ \mathrm{f}\ldots)$.
Note that these correlation functions are eventually independent of $t$, as shown in \appref{appx:coefficient}.
Substituting the two contributions, \eqref{eq:integrand_1} and \eqref{eq:integrand_2}, into the right-hand side of Eq.~\eqref{eq:neumann_born_markov}, we obtain a Lindblad-type equation,
\begin{align}
  \dv{\rho _ \mathrm{q}}{t} 
  &= 
  -i\comm{\frac{\vb{p} ^ +}{\hbar} \cdot \tensor{S} _ {\mathrm{L}} \cdot \frac{\vb{p} ^ -}{\hbar}}{\rho _ \mathrm{q}}
  +i\comm{\frac{\vb{p} ^ -}{\hbar} \cdot \tensor{S} _ \mathrm{G} ^ \top \cdot \frac{\vb{p} ^ +}{\hbar}}{\rho _ \mathrm{q}}
  \notag \\
  &+\qty(\tensor{\Gamma} _ \mathrm{L} \cdot \frac{\vb{p} ^ -}{\hbar}) \cdot \rho _ \mathrm{q} \frac{\vb{p} ^ +}{\hbar}
  -\frac{1}{2}\acomm{\frac{\vb{p} ^ +}{\hbar} \cdot \tensor{\Gamma} _ \mathrm{L} \cdot \frac{\vb{p} ^ -}{\hbar}}{\rho _ \mathrm{q}},
  \notag \\
  &+\qty(\tensor{\Gamma} _ \mathrm{G} ^ \top \cdot \frac{\vb{p} ^ +}{\hbar}) \cdot \rho _ \mathrm{q} \frac{\vb{p} ^ -}{\hbar}
  -\frac{1}{2}\acomm{\frac{\vb{p} ^ -}{\hbar} \cdot\tensor{\Gamma} _ {\mathrm{G}} ^ \top\cdot\frac{\vb{p} ^ +}{\hbar}}{\rho _ \mathrm{q}}
  \label{eq:reduced_ME}
\end{align}
where we defined the non-Hermitian parts of the correlation functions, $\tensor{\Gamma} _ {\mathrm{L}(\mathrm{G})} = \qty(\tensor{\mathcal{C}} _ {\mathrm{L}(\mathrm{G})} - \tensor{\mathcal{C}} _ {\mathrm{L}(\mathrm{G})} ^ \dagger)/(2i)$, and the Hermitian parts, $\tensor{S} _ {\mathrm{L}(\mathrm{G})} = -\qty(\tensor{\mathcal{C}} _ {\mathrm{L}(\mathrm{G})} + \tensor{\mathcal{C}} _ {\mathrm{L}(\mathrm{G})} ^ \dagger)/4$.
As shown in Appendix~\ref{appx:coefficient}, they can be written as
\begin{align}
  &\tensor{\Gamma} _ \mathrm{L} 
  = 2\pi \expval{\vb{E} ^ -(\Rq,\omegaq) \vb{E} ^ +(\Rq,\omegaq)},
  \label{eq: Gamma _ L}
  \\
  &\tensor{\Gamma} _ \mathrm{G}
  = 2\pi \expval{\vb{E} ^ +(\Rq,\omegaq) \vb{E} ^ -(\Rq,\omegaq)} ^ \top,
  \label{eq: Gamma _ G}
  \\
  &\tensor{S} _ \mathrm{L} 
  = \PV\int
  \frac{
    \expval{\vb{E} ^ -(\Rq,\omega) \vb{E} ^ +(\Rq,\omega)}
  }{\omegaq - \omega}
  \dd{\omega},
  \label{eq: S _ L}
  \\
  &\tensor{S} _ \mathrm{G} 
  = \PV\int
  \frac{
    \expval{\vb{E} ^ +(\Rq,\omega) \vb{E} ^ -(\Rq,\omega)} ^ \top
  }{\omegaq - \omega}
  \dd{\omega}.
  \label{eq: S _ G}
\end{align}
While the non-Hermitian parts $\tensor{\Gamma} _ \mathrm{L,G}$ are responsible for irreversible evolution, the Hermitian parts $\tensor{S} _ \mathrm{L,G}$ give an additional unitary evolution of the two-level system due to the interaction with the surrounding environment.
The effect of this additional unitary evolution can be interpreted as a type of Lamb shift, which is typically negligibly small~\cite{breuer2002theory}. 
However, in passive environments it is responsible for the Casimir-Polder force~\cite{casimir1948influence,sukenik1993measurement}, which is \textit{normal} to the surface.
As we will focus on the fluctuation-induced \textit{lateral} forces in the following, the Lamb-type shifts may be neglected.
It is interesting to note that the Lamb shift may provide a clear spectroscopic signature of how the electro-optic interaction renormalizes the atomic response, akin to quantum friction~\cite{klatt2016spectroscopic,durnin2022spectroscopic,franca2025spectroscopic}.
A detailed analysis of this effect is left for future work.

In the second line of Eq.~\eqref{eq:reduced_ME}, we can recognise that the relaxation operator $\vb{p} ^ -$ acts at the left to the reduced density matrix $\rho _ \mathrm{q}$ together with $\tensor{\Gamma} _ \mathrm{L}$.
From this observation, the matrix $\tensor{\Gamma} _ \mathrm{L}$ describes a relaxation process that promotes downward transitions in the two-level system.
A similar argument can be made for $\tensor{\Gamma} _ \mathrm{G}$, which describes an excitation process that promotes upward transitions.
In other words, the correlation function $\expval{\vb{E} ^ - \vb{E} ^ +}$, which is proportional to $\tensor{\Gamma} _ \mathrm{L}$, controls the (spontaneous) photon emission of the two-level system, while the correlator $\expval{\vb{E} ^ + \vb{E} ^ -}$, which is proportional to $\tensor{\Gamma} _ \mathrm{G}$, governs the photon absorption by the two-level system.

These spectroscopic signatures can be discussed in relation to the fluctuation-induced forces~\cite{klatt2016spectroscopic,durnin2022spectroscopic,franca2025spectroscopic}.
In particular, the resulting Lamb shift provides a clear spectroscopic signature of how the electro-optic interaction renormalizes the atomic response. A detailed analysis of this effect is left for future work.

As we shall further discuss in the next section, the fluctuation-induced forces are determined by the steady-state properties of the two-level system, where the system no longer evolves in time.
The steady-state populations can be found from the reduced master equation~\eqref{eq:reduced_ME}.
Since the Lindblad-type terms, which are responsible for the irreversible evolution [i.e., the second and third lines in Eq.~\eqref{eq:reduced_ME}], spoil the off-diagonal elements (coherence) of the density matrix, one may expect that the density matrix becomes diagonal in the steady state ($t \to \infty$).
The commutators in Eq.~\eqref{eq:reduced_ME}, which are responsible for the Lamb-type shifts, vanish if the density matrix is diagonal.
Thus, we may safely neglect the Lamb-type shifts in evaluating the steady-state density matrix $\rho _ \mathrm{q}(\infty)$:
At the steady state, we can set $\dv*{t} = 0$ and solve the equation for $\mel{\ell _ 1}{\rho _ \mathrm{q}(\infty)}{\ell _ 2}\,(\ell _ {1,2} \in \qty{0,1})$ to get
\begin{subequations}
\begin{align}
  &\expval{\rho _ \mathrm{q}(\infty)}{0}
  =\frac{
    \vb{d} _ \mathrm{e} ^ * \cdot \tensor{\Gamma} _ \mathrm{L} \cdot \vb{d} _ \mathrm{e}
  }{
    \vb{d} _ \mathrm{e} ^ * \cdot \tensor{\Gamma} _ {\mathrm{G}}\cdot \vb{d} _ \mathrm{e}
    +\vb{d} _ \mathrm{e} ^ * \cdot \tensor{\Gamma} _ \mathrm{L} \cdot \vb{d} _ \mathrm{e}
  },
  \label{eq:rho _ 00}
  \\
  &\expval{\rho _ \mathrm{q}(\infty)}{1}
  =\frac{
    \vb{d} _ \mathrm{e} ^ * \cdot \tensor{\Gamma} _ \mathrm{G} \cdot \vb{d} _ \mathrm{e}
  }{
    \vb{d} _ \mathrm{e} ^ * \cdot \tensor{\Gamma} _ {\mathrm{G}}\cdot \vb{d} _ \mathrm{e}
    +\vb{d} _ \mathrm{e} ^ * \cdot \tensor{\Gamma} _ \mathrm{L} \cdot \vb{d} _ \mathrm{e}
  }.
  \label{eq:rho _ 11}
\end{align}
Note that we used $\tr\qty{\rho _ \mathrm{q}} = 1$.
It is clear that the ground-state (excited-state) population becomes unity (zero) in the absence of optical gain as it should be.
It can be analytically confirmed that, in the steady state,
\begin{align}
  \mel{0}{\rho _ \mathrm{q}(\infty)}{1} = \mel{1}{\rho _ \mathrm{q}(\infty)}{0} = 0.
  \label{eq:rho-offdiag}
\end{align}
\end{subequations}
Thus, the steady-state density matrix is indeed diagonal, and the steady-state populations are not affected by the Lamb-type shifts, as anticipated.

\section{Fluctuation-induced Hall-like lateral force}
\label{sec:forces}

The fluctuation-induced lateral force can be found from the expectation value of the derivative of the interaction energy with respect to the lateral position $\xq$ of the particle,
\begin{align}
  \vb{F} _ \parallel = 
  \tr{-\rho(t)\partial _ {\xq} H _ \mathrm{int}(t)}.
\end{align}
Here, we first construct a reduced force operator, which is compatible with the Lindblad description of the two-level system.
We use the approximation $\rho(t) \approx \rho(0) + (i\hbar) ^ {-1}\int _ 0 ^ \infty \comm{H _ \mathrm{int}(t-s)}{\rho _ \mathrm{q}(t)\otimes\rho _ \mathrm{f}} \dd{s}$, as we did in deriving the reduced master equation.
We suppose that $\tr\qty{\rho(0)\partial _ {\xq} H _ \mathrm{int}} = 0$ 
and therefore the term $\rho(0)$ can be ignored [note that the interaction Hamiltonian $H _ \mathrm{int}$ contains a single relaxation (or excitation) operator so that taking the trace with the initial density matrix $\rho(0)$ gives a trivial contribution].
After some algebra, the lateral force expectation can be expressed in terms of the density matrix of the two-level system as:
\begin{subequations}
\begin{align}
  &\vb{F} _ \parallel
  =\tr\qty{
    \qty(
    \frac{\vb{p} ^ -}{\hbar} \cdot 
    \qty[\tensor{\mathcal{F}} _ \mathrm{G} ^ \parallel] ^ \top
    \cdot \frac{\vb{p} ^ +}{\hbar}
    -\frac{\vb{p} ^ +}{\hbar} \cdot \tensor{\mathcal{F}} _ \mathrm{L} ^ \parallel \cdot \frac{\vb{p} ^ -}{\hbar}
    )
    \rho _ \mathrm{q}(t)
  }.
  \label{eq:F _ Born-Markov}
\end{align}
Here $\tensor{\mathcal{F}} _ \mathrm{G(L)}$ are matrices that determine the momentum transfer in the gain (loss) channel,
\begin{align}
  &\tensor{\mathcal{F}} _ \mathrm{G} ^ \parallel =
  \qty[
  \qty{-i\hbar \partial _ {\xq}}
  2\pi 
  \expval{
    \vb{E} ^ +(\Rq', \omegaq) 
    \vb{E} ^ -(\Rq, \omegaq)
  } ^ \top 
  ] _ {\Rq' = \Rq},
  \label{eq:F _ G}
  \\
  &\tensor{\mathcal{F}} _ \mathrm{L} ^ \parallel =
  \qty[
  \qty{-i\hbar \partial _ {\xq}}
  2\pi 
  \expval{
    \vb{E} ^ -(\Rq, \omegaq) 
    \vb{E} ^ +(\Rq', \omegaq)
  }
  ] _ {\Rq' = \Rq}.
  \label{eq:F _ L}
\end{align}
\end{subequations}
Recall that we defined $\Rq = \xq + \zq\uz$.
It is also useful to mention that the normal component of the force contains additional terms related to the principal-value integrals $S _ \mathrm{G,L}$, which represent the Lamb-type shifts (see, e.g., Ref.~\cite{buhmann2004casimir, hassani2018spontaneous, silveirinha2018fluctuation}).
Our result for the lateral force agrees with the previous literature~\cite{hassani2018spontaneous, silveirinha2018fluctuation}, when the material is passive.
In the passive case, the lateral force always vanishes when the two-level system is in the ground state (i.e., $\expval{\rho _ \mathrm{q}}{0}=1$). 
In contrast, in systems with the chiral gain, the lateral force can remain finite even if $\expval{\rho _ \mathrm{q}}{0}=1$.

We can regard the quantity inside the curved parentheses in Eq.~\eqref{eq:F _ Born-Markov} as the reduced force operator.
The first (second) term within the parentheses has the excitation (relaxation) operator $\vb{p} ^ {+(-)}$ on the far right, which acts on the density matrix, so that we can understand that it corresponds to the gain (loss) contribution.
The operator in the curly brackets in Eqs.~\eqref{eq:F _ G} and \eqref{eq:F _ L} corresponds to the lateral component of the canonical momentum operator of the electromagnetic field. 
As the field correlators can be written in terms of the momentum spectra [see Eqs.~\eqref{eq:gamma _ L} and \eqref{eq:gamma _ G}], the expression can be simplified in the reciprocal space,
\begin{align}
\tensor{\mathcal{F}} _ \mathrm{G,L} =
\int \hbar\vb{k}\,
2\pi\tensor{\gamma} _ \mathrm{G,L} (\vb{k},\zq)
\dd{\vb{k}}.
\label{eq:F _ G,L (reciprocal)}
\end{align}
Recall that we defined $\dd{\vb{k}} := \dd{k _ x}\dd{k _ y}/(2\pi) ^ 2.$
The expression \eqref{eq:F _ G,L (reciprocal)} shows that the force can be expressed as the integral of the transferred momentum $\hbar\vb{k}$ multiplied by the spectra $\tensor{\gamma} _ \mathrm{G,L}$ of photon absorption and emission in each channel.
The matrices $\tensor{\mathcal{F}} _ \mathrm{G,L}$ represent the momentum transfer rates in the respective processes.

As illustrated in FIGs.~\ref{fig:E-E+} and \ref{fig:E+E-}, the matrices $\tensor{\gamma} _ \mathrm{G,L}$ are even functions of $k _ x$.
This implies that the overall integrands in Eq.\,\eqref{eq:F _ G,L (reciprocal)} are odd functions of $k _ x$, and thereby the $x$ component $F _ x = \ux \cdot \vb{F} _ \parallel$ of the force vanishes.
This result is somewhat counterintuitive: as the electric current induced by the bias field flows along x, one would normally expect a frictional drag opposing the carrier motion.
Surprisingly, in our system the lateral force is instead \textit{perpendicular} to the bias field, resembling a Hall-like effect.

\begin{figure}[tbp]
  \centering
  \includegraphics[width=.9\linewidth]{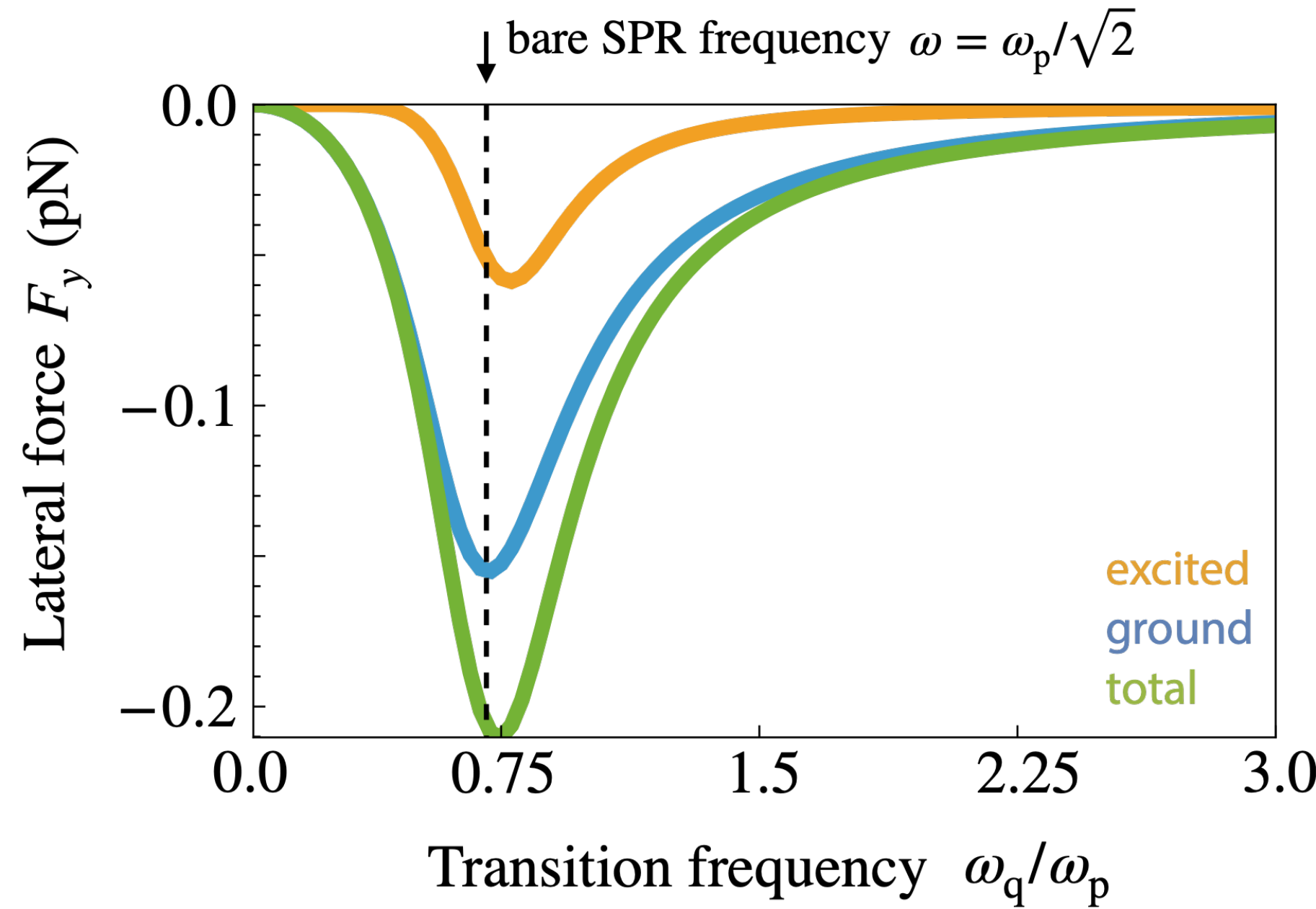}
  \caption{
    Fluctuation-induced lateral force $F _ y = \uy \cdot \vb{F} _ \parallel$ as a function of the transition frequency of the two-level system.
    The green curve represents the total lateral force.
    The orange (blue) curve corresponds to the contribution from the excited (ground) state, corresponding to Eqs.~\eqref{eq:F _ L} and \eqref{eq:F _ G}, respectively.
    The force is exerted along the negative $y$ direction.
    The following parameters were used to generate the plots:
    $\vb{d} _ \mathrm{e}/\abs{\vb{d} _ \mathrm{e}} = \uz$,
    $\abs{\vb{d} _ \mathrm{e}} = 100\,\mathrm{D}$,
    $\omega _ 0/\omegap = 0.1$,
    $\gamma/\omegap = 0.5$,
    $\omegap z _ \mathrm{q}/c = 0.0001$,
    and
    $\omegap/(2\pi) = 1.0\,\mathrm{THz}$.
    The dashed line represents the surface plasmon resonance frequency in the absence of chiral gain ($\omega \approx \omegap/\sqrt{2}$).
  }
  \label{fig:F}
\end{figure}
In \figref{fig:F}, we depict the fluctuation-induced lateral force $F _ y = \uy \cdot \vb{F} _ \parallel$ as a function of the transition frequency $\omegaq$ of the two-level system.
The parameters of the chiral-gain material assumed for the figure are analogous (but not exactly coincident) to those of the electro-optical response of p-doped trigonal tellurium under a bias strength on the order of $3 \times 10^5\ \mathrm{V/m}$~\cite{tsirkin2018gyrotropic,morgado2024non}: the effective plasma frequency of tellurium (permittivity near zero point) is given as $\omegap ^ \mathrm{eff}/(2\pi) \approx 1.2\ \mathrm{THz}$, and the damping constant is $\gamma \approx 1.6 \times 10 ^ {12}\ \mathrm{rad/s}$~\cite{morgado2024non}; hence, we can estimate $\gamma/\omegap ^ \mathrm{eff} \approx 0.2$ and $\omega _ 0/\omegap ^ \mathrm{eff} \approx 0.1$ for $E _ \mathrm{bias} = 3 \times 10^5\ \mathrm{V/m}$.
In the calculation, we use the ``equilibrium'' density matrix [$\rho _ \mathrm{q}(t) \rightarrow \rho _ \mathrm{q}(\infty)$], as the two-level system eventually relaxes to the steady state.
The fluctuation-induced lateral force is exerted due to the chiral optical gain and is significantly enhanced near the surface plasmon resonance condition $\omega = \omegap/\sqrt{2}$.
The peak slightly deviates from the `bare' resonance condition due to the NHEO effect.
The lateral force is along the negative $y$ direction.

This is consistent with the following qualitative discussion.
As we have seen in the momentum spectra representing the photon emission (absorption), the two-level system is more likely to emit (absorb) photons propagating in the positive (negative) $y$ direction.
Thus, the recoil force generated during photon emission is directed along the negative $y$-direction.
Similarly, the `kick' imparted to the two-level system during photon absorption is also directed in the negative $y$-direction.
Consequently, the net force acting on the two-level system points towards the negative $y$ direction, in agreement with the numerical simulations.
It is also worth noting that the $+y$-direction corresponds to the propagation direction of plasmons that experience gain due to the NHEO effect.
The emission of such plasmons is stimulated by the two-level system, which in turn experiences a corresponding recoil force.

\begin{figure}[tbp]
  \centering
  \includegraphics[width=.85\linewidth]{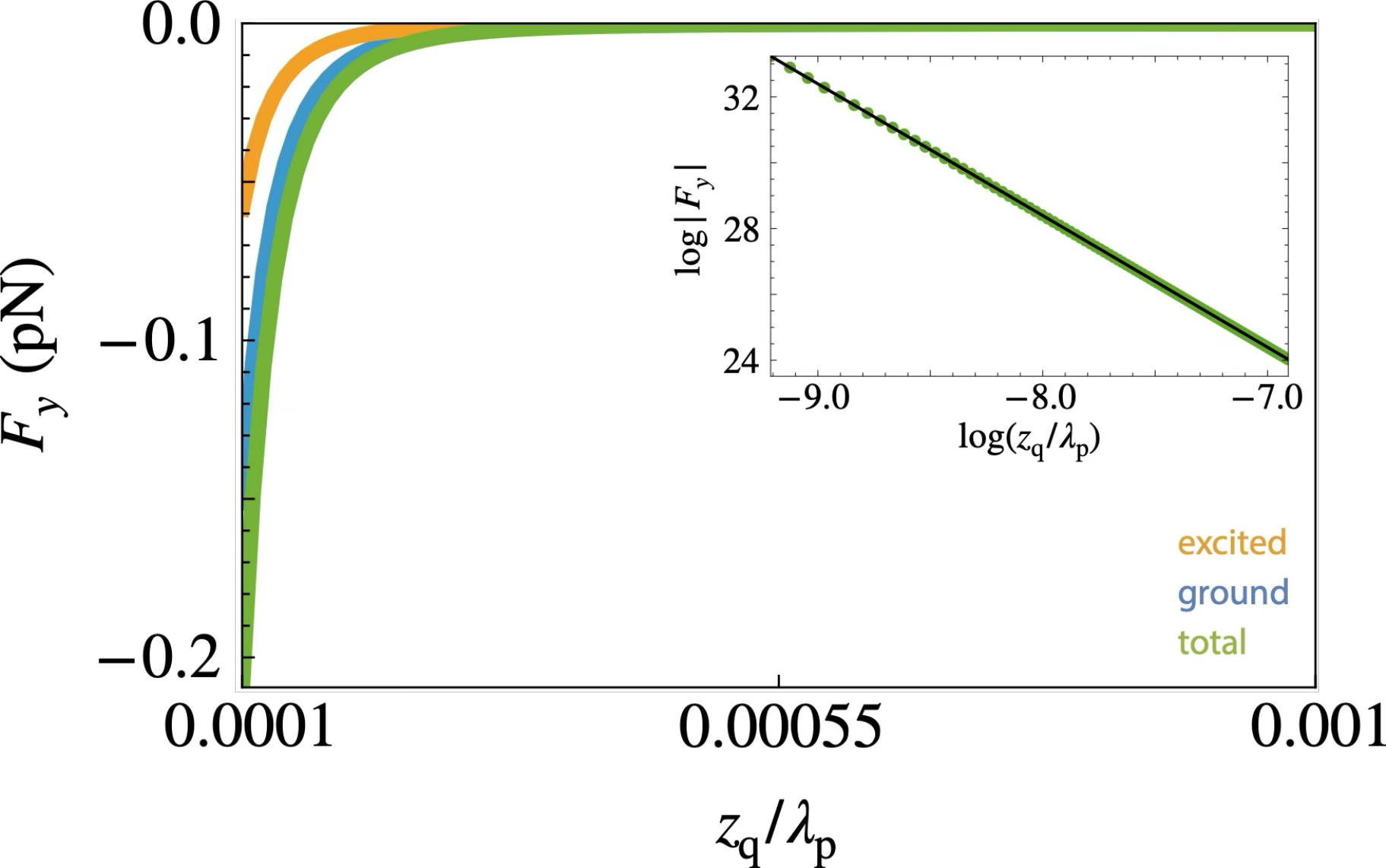}
  \caption{
    The lateral force as a function of the distance between the two-level system and the surface.
    The inset is a double logarithmic plot.
    The fitting of the logarithmic plot indicates that the force scales with the distance as $1/z _ \mathrm{q} ^ 4$.
    We considered $\omegaq/\omegap = 0.73$; the other parameters used to generate the plot are the same as those of \figref{fig:F}.
    The distance $\zq$ is normalised by the plasma wavelength $\lambda _ \mathrm{p} = c/\omegap$.
  }
  \label{fig:Fy-zq}
\end{figure}

It can be verified that the lateral force scales with the distance to the slab approximately as $1/z _ \mathrm{q} ^ 4$ (see \figref{fig:Fy-zq}).
This scaling behaviour is the same as that of the standard (vertical) Casimir force near a metallic slab when retardation effects are negligible.

As previously noted, the lateral force is perpendicular to the expected direction of frictional drag arising from carrier motion in the conductor.
It may be interpreted as a Hall-like force rooted in the quantum geometry of the low-symmetry conductor.
Consistent with this picture, the sign of the lateral force reverses when the electric-bias direction is switched.
 Furthermore, its magnitude is proportional to the cyclotron-type frequency $\omega _ 0$, and is therefore directly governed by the Berry-curvature dipole $D$.
 Thus, much like the Berry dipole can give rise to nonlinear Hall effects~\cite{sodemann2015quantum,du2021nonlinear}, our analysis shows that it can also generate Hall-like fluctuation forces.

To conclude, we note that in principle, a force component along the direction of motion may also exist.
However, such a contribution typically depends on Doppler shifts in the material response due to particle motion, which are negligible in our calculation~\cite{shapiro2010thermal,volokitin2011quantum,dean2016nonequilibrium,shapiro2017fluctuation,oue2025quantum} (see Appendix \ref{appx:doppler} for more details).

\section{Conclusion}
\label{sec:conclusion}
In this study, we have uncovered fluctuation-induced lateral forces acting on a small particle which is modelled by a two-level system and placed near a translation invariant gain medium substrate.
Unlike in passive environments, we found that the two-level system experiences a persistent lateral force even when it is in the ground state.
Moreover, the sign of the force is independent of the atomic state.
This suggests that gain-assisted environments can be used to induce and control the lateral motion of small particles, opening up intriguing possibilities for optical manipulation based on fluctuation-induced fields.

In our system, the gain arises due to the non-Hermitian electro-optic effect, which produces a chiral gain response.
We demonstrated how this chiral gain modifies the propagation of surface plasmon polaritons in the material and, in turn, shapes the momentum spectra of emission and absorption processes.

The momentum transfer associated with its relaxation and excitation processes produces a lateral force on the two-level system.
Specifically, our analysis shows that the momentum transfer rates associated with the relaxation and excitation processes are described by two distinct field-correlation functions, which can exhibit pronounced asymmetry in the direction perpendicular to the applied electric bias with the spectral asymmetry governed by the Berry curvature dipole.
Consequently, the lateral force can be interpreted as a Hall-like fluctuation-induced force, establishing a novel and exciting connection between the quantum geometry of the material and its manifestations in fluctuation electrodynamics.

\begin{acknowledgments}
  D.O.~is supported by JSPS Overseas Research Fellowship and by the RIKEN special postdoctoral researcher program. M.S. is partially supported by the Simons Foundation (award SFI-MPS-EWP-00008530-10) and by national funds through FCT -- Funda\c{c}\~{a}o para a Ci\^{e}ncia e a Tecnologia, I.P., and, when eligible, co-funded by EU funds under project/support UID/50008/2025 -- Instituto de Telecomunica\c{c}\~{o}es, with DOI identifier \url{https://doi.org/10.54499/UID/50008/2025}.
\end{acknowledgments}

\section*{Data availability}
The data that support the findings of this article are not publicly available.
The data are available upon reasonable request from the authors.

\appendix
\section{quasistatic Green's function}
\label{appx:green}
In the quasistatic regime, electric and magnetic fields are decoupled, and we can write
\begin{align}
  &\vb{E} = -\nabla \phi,
  \label{eq:E_phi}
  \\
  &\nabla \times \vb{E} = 0,
  \label{eq:faraday_quasistatic}
  \\
  &\nabla \cdot \tensor\epsilon \cdot \vb{E} = \frac{\rho}{\epsilon _ 0} 
  = \frac{-1}{\omega ^ 2/c ^ 2}\nabla \cdot i\omega\mu _ 0 \vb{j}.
  \label{eq:gauss}
\end{align}
where we introduced the electrostatic potential $\phi$ and the electric charge density $\rho$ and used the continuity equation $-i\omega \rho + \nabla \cdot \vb{j} = 0$.
The quasi-static approximation holds in the near-field region, where electromagnetic propagation delays are negligible and the speed of light can be effectively treated as infinite.

Since Eq.~\eqref{eq:E_phi} automatically satisfies  Eq.~\eqref{eq:faraday_quasistatic}, we can focus on the third equation~\eqref{eq:gauss}.
Substituting Eq.~\eqref{eq:E_phi} into Eq.~\eqref{eq:gauss}, we can get
\begin{align}
    -\nabla ^ 2 \phi = \frac{-1}{\omega ^ 2/c ^ 2}\frac{1}{\varepsd}\nabla \cdot i\omega \mu _ 0\vb{j}.
\end{align}
Note that we have $\tensor\epsilon = \epsilon _ \mathrm{d}I _ {3\times3} + i\epsilon _ \mathrm{g} \vb{u} _ x \times I _ {3\times3}$ for $z < 0$ (and $\tensor\epsilon = I _ {3\times3}$ for $z > 0$); thus, we can write $\nabla \cdot \tensor\epsilon \cdot \nabla = \varepsilon _ \mathrm{d} \nabla ^ 2$ with $\varepsd (z) = \theta(+z) + \epsd \theta(-z)$.

We introduce a scalar Green's function, which is defined as
\begin{align}
  &g(\vb{r},\vb{r}') =
  \int g _ {\vb{k}} (z,z') e ^ {i\vb{k}\cdot\qty(\vb{x}-\vb{x}')}
  \dd{\vb{k}},
  \\
  &(\vb{k} ^ 2 - \partial _ {z} ^ 2) g _ {\vb{k}}(z,z') = \delta _ {z,z'},
  \label{eq:poisson_FT}
\end{align}
where we defined $\vb{k} = k _ x \vb{u} _ x + k _ y \vb{u} _ y$.
Note that we used the shorthand notation $\dd{\vb{k}} = \dd{k _ x}\dd{k _ y}/(2\pi) ^ 2$.
Then, the electric field can be written as
\begin{align}
  &\vb{E}(\vb{r})
  = -\nabla \int 
  g(\vb{r},\vb{r}')
  \frac{-1}{\omega ^ 2/c ^ 2}
  \frac{1}{\varepsd(z')}
  \nabla' \cdot i\omega \mu _ 0 \vb{j}(\vb{r}') \dd{\vb{r}'}, 
  \nonumber
  \\
  &= \int 
  \qty{\frac{-1}{\omega ^ 2/c ^ 2}
  \nabla \nabla' \frac{g(\vb{r},\vb{r}')}{\varepsd(z')}}
  \cdot
  i\omega\mu _ 0 \vb{j}(\vb{r}')\dd{\vb{r}'},
\end{align}
where $\nabla'$ represents the derivative with respect to $\vb{r}'$.
Note that we performed the integration by parts.
Comparing this equation with Eq.~\eqref{eq:E=Gj}, we find that the Green's function can be expressed as:
\begin{align}
  \tensor{G}(\vb{r},\vb{r}') =
  \frac{-1}{\omega ^ 2/c ^ 2}
  \nabla \nabla'
  \frac{g(\vb{r},\vb{r}')}{\varepsd(z')}.
\end{align}

Solving the Fourier-transformed Laplace equation $\qty(\vb{k} ^ 2 - \partial _ z ^ 2) \phi _ {\vb{k}}(z) = 0$ above and below the surface ($z = 0$) and imposing the field continuity conditions at the surface, we can write $g(\vb{r}, \vb{r}')$ with the help of transmission ($t _ {\vb{k}}$) and reflection ($r _ {\vb{k}}$) coefficients for plane-wave incidence from the chiral-gain medium to the air region.
These coefficients are explicitly calculated in Appendix~\ref{appx:transmission}. 
In the relevant spatial region, the scalar Green's function is:
\begin{align}
  g _ {\vb{k}}(z,z') = 
  \frac{t _ {\vb{k}}}{2\abs{\vb{k}}}
  e ^ {-\abs{\vb{k}}(z - z')}
  \quad
  (z' < 0 < z),
\end{align}
Finally, the Green's function dyadic can be expressed as, 
$
\tensor{G}(\vb{r},\vb{r}') =
\int
\tensor{G} _ {\vb{k}}(z,z')
e ^ {i\vb{k}\cdot\qty(\vb{x}-\vb{x}')}
\dd{\vb{k}},
$
with the Fourier amplitude given by (for $z' < 0 < z$):
\begin{align}
  \tensor{G} _ {\vb{k}}(z,z') =
  \frac{-1}{\omega ^ 2/c ^ 2}
  \frac{t _ {\vb{k}}/\epsd}{2\abs{\vb{k}}}
  \vb{k} _ + \vb{k} _ +
  e ^ {-\abs{\vb{k}}(z - z')},
\end{align}
where we defined $\vb{k} _ + = \vb{k} + i\abs{\vb{k}}\uz$.

\section{Transmission coefficient}
\label{appx:transmission}
To obtain the scalar Green's function introduced in Appendix \ref{appx:green}, we solve $-\nabla \cdot \tensor{\epsilon} \cdot \nabla \phi = 0$, above and below the interface,
\begin{align}
  \phi _ {\vb{k}} (z) =
  \begin{cases}
  t  _ {\vb{k}} e ^ {-\abs{\vb{k}}z}
  &(z > 0),
  \\
  e ^ {-\abs{\vb{k}}z} + r  _ {\vb{k}} e ^ {+\abs{\vb{k}}z}
  &(z < 0),
  \end{cases}
\end{align}
where we assumed an incoming ``plane wave''  propagating towards $+z$ in the chiral-gain medium.
Here, $t  _ {\vb{k}}$ and $r  _ {\vb{k}}$ are the transmission and reflection coefficients, respectively.
The relevant boundary conditions are derived from the continuity of the electrostatic potential and the continuity of the electric displacement field:
\begin{align}
  &\phi(z = 0 ^ +) - \phi(z = 0 ^ -) = 0,
  \label{eq:phi_continuous_after}
  \\
  &\eval{\pdv{\phi}{z}} _ {z = 0 ^ +} - \epsilon _ \mathrm{d} \eval{\pdv{\phi}{z}} _ {z = 0 ^ -} = i\epsilon _ \mathrm{g} \eval{\pdv{\phi}{y}} _ {z = 0 ^ -},
  \label{eq:phi'_discontinuous_after}
\end{align}
where $0 ^ {+(-)}$ is the positive (negative) infinitesimal.
Imposing these boundary conditions, we obtain the following explicit formulas for the reflection and transmission coefficients:
\begin{align}
  r  _ {\vb{k}} = \frac{\epsilon _ \mathrm{d}\abs{\vb{k}} + \epsilon _ \mathrm{g}k _ y - \abs{\vb{k}}}{\epsilon _ \mathrm{d}\abs{\vb{k}} - \epsilon _ \mathrm{g}k _ y + \abs{\vb{k}}},
  \quad
  t  _ {\vb{k}} = \frac{2\epsilon _ \mathrm{d}\abs{\vb{k}}}{\epsilon _ \mathrm{d}\abs{\vb{k}} - \epsilon _ \mathrm{g}k _ y + \abs{\vb{k}}}.
\end{align}

\section{Coefficient matrices in the reduced master equation}
\label{appx:coefficient}
In the reduced master equation~\eqref{eq:reduced_ME}, there are two types of coefficient matrices,
\begin{align}
  \tensor{\Gamma} _ \mathrm{L,G} =
  \frac{1}{2i}\qty(
  \tensor{\mathcal{C}} _ \mathrm{L,G} - \tensor{\mathcal{C}} _ \mathrm{L,G} ^ \dagger
  ),
  \quad
  \tensor{S} _ \mathrm{L,G} = -\frac{1}{4} \qty(
  \tensor{\mathcal{C}} _ \mathrm{L,G} + \tensor{\mathcal{C}} _ \mathrm{L,G} ^ \dagger
  ),
\end{align}
where $\tensor{\mathcal{C}} _ \mathrm{L,G}$ are defined as
\begin{align}
  &\tensor{\mathcal{C}} _ \mathrm{L} =
  2i\int _ 0 ^ \infty
  \expval{\vb{E} ^ -(t) \vb{E} ^ +(t-s)} 
  e ^ {+i\omegaq s}
  \dd{s},
  \label{eq:calG _ L(appx)}
  \\
  &\tensor{\mathcal{C}} _ \mathrm{G} =
  2i\int _ 0 ^ \infty 
  \expval{\vb{E} ^ +(t-s) \vb{E} ^ -(t)} ^ \top
  e ^ {+i\omegaq s}
  \dd{s}.
  \label{eq:calG _ G(appx)}
\end{align}
In this appendix, we obtain simplified formulas for $\tensor{\Gamma} _ \mathrm{L,G}$ and $\tensor{S} _ \mathrm{L,G}$.

We write explicitly the time-dependent field operators as,
\begin{align}
  &\vb{E} ^ -(t) = 
  \int
  \vb{E} ^ -(\omega)
  e ^ {-i\omega t}
  \dd{\omega},
  \label{eq:E-(t)(appx)}
\end{align}
where we applied the shorthand notation $\vb{E} ^ -(\omega) := \vb{E} ^ -(\Rq,\omega)$.
Then, Eq.~\eqref{eq:calG _ L(appx)} can be reduced to
\begin{align}
  &\int _ 0 ^ \infty
  \iint
  \expval{\vb{E} ^ -(\omega) \vb{E} ^ +(\omega')} 
  e ^ {-i(\omega - \omega')t +i(\omegaq - \omega')s}
  \dd{\omega}\dd{\omega'}
  \dd{s}  \nonumber
  \\
  &=\int _ 0 ^ \infty
  \int
  \expval{\vb{E} ^ -(\omega) \vb{E} ^ +(\omega)} 
  e ^ {+i(\omegaq - \omega)s}
  \dd{\omega}
  \dd{s}.
\end{align}
Note that only the equal-frequency part ($\omega' = \omega$) can contribute to the integral so that the integration over $\omega'$ is removed (and we set $\omega' = \omega$ in the integrand).
Similarly, Eq.~\eqref{eq:calG _ G(appx)} becomes
\begin{align}
  \int _ 0 ^ \infty
  \int
  \expval{\vb{E} ^ +(\omega) \vb{E} ^ -(\omega)} 
  e ^ {+i(\omegaq - \omega)s}
  \dd{\omega}
  \dd{s}.
\end{align}
From these expressions, it is clear that the correlation functions $\tensor{\mathcal{C}} _ \mathrm{L,G}$ are independent of time $t$.

The $\tensor{\Gamma} _ \mathrm{L}$ matrix can be written as
\begin{align}
  &\vb{u} _ {\ell _ 1} \cdot \tensor{\Gamma} _ \mathrm{L} \cdot \vb{u} _ {\ell _ 2}   \nonumber
  \\
  &=\int _ 0 ^ \infty
  \int
  \expval{E _ {\ell _ 1} ^ -(\omega) E _ {\ell _ 2}   ^ +(\omega)}
  e ^ {+i(\omegaq - \omega)s}
  \dd{\omega}
  \dd{s}
  \notag \\
  &+
  \int _ 0 ^ \infty
  \int
  \expval{E _ {\ell _ 2} ^ -(\omega) E _ {\ell _ 1} ^ +(\omega)} ^ *
  e ^ {-i(\omegaq - \omega)s}
  \dd{\omega}
  \dd{s} \nonumber
  \\
  &=\int _ 0 ^ \infty
  \int
  \expval{E _ {\ell _ 1} ^ -(\omega) E _ {\ell _ 2} ^ +(\omega)}
  e ^ {+i(\omegaq - \omega)s}
  \dd{\omega}
  \dd{s}
  \notag \\
  &+
  \int _ {-\infty} ^ 0
  \int
  \expval{E _ {\ell _ 2} ^ -(\omega) E _ {\ell _ 1} ^ +(\omega)} ^ *
  e ^ {+i(\omegaq - \omega)s}
  \dd{\omega}
  \dd{s} \nonumber
  \\
  &=\int _ 0 ^ \infty
  \int
  \expval{E _ {\ell _ 1} ^ -(\omega) E _ {\ell _ 2} ^ +(\omega)}
  e ^ {+i(\omegaq - \omega)s}
  \dd{\omega}
  \dd{s}
  \notag \nonumber\\
  &+
  \int _ {-\infty} ^ 0
  \int
  \expval{E _ {\ell _ 1} ^ +(\omega) E _ {\ell _ 2} ^ -(\omega)}
  e ^ {+i(\omegaq - \omega)s}
  \dd{\omega}
  \dd{s} \nonumber
  \\
  &=\int _ {-\infty} ^ \infty
  \int
  \vb{u} _ {\ell _ 1} \cdot
  \expval{\vb{E} ^ -(\omega) \vb{E} ^ +(\omega)}
  \cdot \vb{u} _ {\ell _ 2}  
  e ^ {+i(\omegaq - \omega)s}
  \dd{\omega}
  \dd{s} \nonumber
  \\
  &=
  \vb{u} _ {\ell _ 1} \cdot
  2\pi
  \expval{\vb{E} ^ -(\omegaq) \vb{E} ^ +(\omegaq)}
  \cdot \vb{u} _ {\ell _ 2}.
\end{align}
A similar operation may be applied to evaluate the $\tensor{\Gamma} _ \mathrm{G}$ matrix.
Overall, we have
\begin{align}
  &\tensor{\Gamma} _ \mathrm{L} 
  = 2\pi \expval{\vb{E} ^ -(\Rq,\omegaq) \vb{E} ^ +(\Rq,\omegaq)},
  \label{eq: Gamma _ L(appx)}
  \\
  &\tensor{\Gamma} _ \mathrm{G}
  = 2\pi \expval{\vb{E} ^ +(\Rq,\omegaq) \vb{E} ^ -(\Rq,\omegaq)} ^ \top.
  \label{eq: Gamma _ G(appx)}
\end{align}

On the other hand, the $\tensor{S} _ \mathrm{L}$ matrix can be written as
\begin{align}
  2i \tensor{S} _ \mathrm{L}
  &=\int _ 0 ^ \infty
  \int
  \expval{\vb{E} ^ -(\omega) \vb{E} ^ +(\omega)}
  e ^ {+i(\omegaq - \omega)s}
  \dd{\omega}
  \dd{s}
  \notag \\
  &-
  \int _ 0 ^ \infty
  \int
  \expval{\vb{E} ^ -(\omega) \vb{E} ^ +(\omega)}
  e ^ {-i(\omegaq - \omega)s}
  \dd{\omega}
  \dd{s} \nonumber
  \\
  &=i\int\qty(
  \frac{
    \expval{\vb{E} ^ -(\omega) \vb{E} ^ +(\omega)}
  }{\omegaq - \omega + i0 ^ +}
  +\frac{
    \expval{\vb{E} ^ -(\omega) \vb{E} ^ +(\omega)}
  }{\omegaq - \omega - i0 ^ +}
  ) \dd{\omega} \nonumber
  \\
  &=2i\PV\int
  \frac{
    \expval{\vb{E} ^ -(\omega) \vb{E} ^ +(\omega)}
  }{\omegaq - \omega}
  \dd{\omega},
\end{align}
where $\PV$ stands for Cauchy's principal value.
A similar treatment gives an analogous expression for $\tensor{S} _ \mathrm{G}$.
In summary, we have
\begin{align}
  \tensor{S} _ \mathrm{L} 
  &= \PV\int
  \frac{
    \expval{\vb{E} ^ -(\Rq,\omega) \vb{E} ^ +(\Rq,\omega)}
  }{\omegaq - \omega}
  \dd{\omega},
  \label{eq: S _ L(appx)}
  \\
  \tensor{S} _ \mathrm{G} 
  &= \PV\int
  \frac{
    \expval{\vb{E} ^ +(\Rq,\omega) \vb{E} ^ -(\Rq,\omega)} ^ \top
  }{\omegaq - \omega}
  \dd{\omega}.
  \label{eq: S _ G(appx)}
\end{align}

\section{Doppler contribution}
\label{appx:doppler}
In our system, the carrier drift can also modify the permittivity through a Doppler effect~\cite{hu1991two}.
Specifically, the Drude component is altered as
$\epsd(\omega) \to \epsd(\omega - k _ x v _ \mathrm{drift})$.
The influence is characterised by the Doppler shift $\Delta = k _ x v _ \mathrm{drift}$; thus, it may, in principle, contribute to the force along the $x$-direction, but not along the perpendicular ($y$-) direction, which is the main focus of this study.

As clarified in our previous works~\cite{morgado2018drift,morgado2021active,oue2024stable,oue2025quantum}, the Doppler shift becomes relevant only in the deep subwavelength regime ($k _ x \zq \gg 1$) with extremely large drift velocities.
However, the contribution from such deep subwavelength modes to the force is exponentially suppressed by the factor $\exp(-2\abs{\vb{k}}\zq)$. 
In contrast, the Berry-dipole contribution is independent of $k _ x$ and therefore much less affected by this suppression.

To be more quantitative, next, we compare the Doppler and Berry-dipole contributions to the dielectric response. Up to first order in the drift velocity, the Doppler contribution to the permittivity is
\begin{align}
  \epsilon _ \mathrm{Doppler} 
  := \epsd(\omega - k _ x v _ \mathrm{drift}) - \epsd(\omega)
  \approx
  -k _ x v _ \mathrm{drift} \dv{\epsilon _ \mathrm{d}}{\omega},
\end{align}
where $\epsilon _ \mathrm{d} = 1 - \omega _ \mathrm{p} ^ 2 / (\omega ^ 2 + i\omega\gamma)$, and the drift velocity is $v _ \mathrm{drift} = e E _ \mathrm{bias}/(m _ \mathrm{e} \gamma)$ with the electron mass $m _ \mathrm{e}$.
Recall that the Berry-dipole contribution is given by Eq.~\eqref{eq: epsg}.
With the parameters used in the calculation of the lateral force ($D = 10 ^ {-4}$ and $\gamma = 0.5\omega _ \mathrm{p}$), the ratio between the two contributions around the surface plasmon resonance frequency is estimated as
\begin{align}
  \abs{\frac{\epsilon _ \mathrm{Doppler}}{\epsilon _ \mathrm{g}}} _ {\omega = {\omega _ \mathrm{p}}/{\sqrt{2}}} 
  \approx 5.0 \times 10 ^ {-8} \abs{\frac{k _ x}{k _ \mathrm{p}}},
  \notag
\end{align}
where the plasma wavenumber is denoted by $k _ \mathrm{p} = \omega _ \mathrm{p}/c$.
Note that both $\epsilon _ \mathrm{Doppler}$ and $\epsg$ are proportional to the bias $E _ \mathrm{bias}$ so that the ratio is independent of the bias.
Hence, only the deep subwavelength components of the Doppler contribution could be comparable to the Berry dipole contribution, but these are exponentially suppressed, and the expected Doppler-induced forces should be very small.
Therefore, the Doppler effect can be safely neglected in the present regime.

\bibliography{mini}
\end{document}